\documentclass[fleqn,usenatbib]{mnras}

\usepackage{newtxtext,newtxmath}

\usepackage[T1]{fontenc}

\DeclareRobustCommand{\VAN}[3]{#2}
\let\VANthebibliography\thebibliography
\def\thebibliography{\DeclareRobustCommand{\VAN}[3]{##3}\VANthebibliography}

\DeclareRobustCommand{\DE}[3]{#2}
\let\DEthebibliography\thebibliography
\def\thebibliography{\DeclareRobustCommand{\DE}[3]{##3}\DEthebibliography}

\usepackage{graphicx}	
\usepackage{amsmath}	
\usepackage{caption}
\usepackage{subcaption}
\usepackage{dcolumn}
\usepackage{array,booktabs}

\newcommand{\msun}{\ensuremath{\mbox{M}_{\odot}}}
\newcommand{\lsun}{\ensuremath{\mbox{L}_{\odot}}}
\newcommand{\rsun}{\ensuremath{\mbox{R}_{\odot}}}

\newcommand{\teff}{\ensuremath{T_{\rm eff}}}
\newcommand{\dHp}{\ensuremath{ \Delta{H}_{\rm P} }}
\newcommand{\drbydt}{ \ensuremath{{\rm d}\rho/{\rm d}{t}}   }
\newcommand{\dtbydt}{\ensuremath{{\rm d}\theta_{\rm B}/{\rm d}{t}}}
\newcommand\T{\rule{0pt}{2.6ex}}       
\newcommand\B{\rule[-1.2ex]{0pt}{0pt}} 

\newcommand{\loggp}{\ensuremath{\log(g_{\rm p})}}

\newcommand{\vsini}{\ensuremath{v_{\rm e}\sin{i}}}

\title[Rotation of $\theta$~Sco~A]{A study of the F-giant star $\theta$~Scorpii~A:  a post-merger rapid rotator?}

\author[Fiona Lewis et al.]{Fiona Lewis$^{1}$\thanks{E-mail: lewis.fiona@gmail.com},
Jeremy Bailey$^{1}$\thanks{E-mail: j.bailey@unsw.edu.au},
Daniel V. Cotton$^{2,3,4,5}$,
Ian D. Howarth$^{6}$,
Lucyna Kedziora-Chudczer$^{5}$,
\newauthor Floor van Leeuwen$^{7}$
\\
\\
$^{1}$School of Physics, University of New South Wales, Sydney, NSW 2052, Australia\\
$^{2}$Monterey Institute for Research in Astronomy, 200 Eighth Street, Marina, CA, 93933, USA.\\
$^{3}$Anglo Australian Telescope, Australian National University, 418 Observatory Road, Coonabarabran, NSW 2357, Australia.\\
$^{4}$Western Sydney University, Locked Bag 1797, Penrith-South DC, NSW 1797, Australia.\\
$^{5}$University of Southern Queensland, Centre for Astrophysics, Toowoomba, QLD 4350, Australia\\
$^{6}$University College London, Gower Street, London WC1E 6BT, UK.\\
$^{7}$Institute of Astronomy, University of Cambridge, Madingley Road, Cambridge CB3 0HA, UK.\\
}

\date{Accepted 2022 April 4. Received March 27; in original form 2022 January 30}

\pubyear{2022}

\begin{document}
\label{firstpage}
\pagerange{\pageref{firstpage}--\pageref{lastpage}}
\maketitle

\begin{abstract}
We report high-precision observations of the linear polarization of the F1$\,$III star $\theta$~Scorpii.  The polarization has a wavelength dependence of the form expected for a rapid rotator, but with an amplitude several times larger than seen in otherwise similar main-sequence stars. 
This confirms the expectation that lower-gravity
stars should have stronger rotational-polarization signatures as a consequence of the density dependence of the ratio of scattering to absorption opacities. 
By modelling the polarization, together with additional observational constraints (incorporating a revised analysis of \textit{Hipparcos} astrometry, which clarifies the system's binary status), we determine a set of precise stellar parameters, including a
rotation rate 
$\omega\, (= \Omega/\Omega_{\rm c})\ge 0.94$,  
polar gravity $\loggp = 2.091 ^{+0.042}_{-0.039}$ (dex cgs),  
mass  $3.10 ^{+0.37}_{-0.32}$ \msun, and 
luminosity $\log(L/\lsun) =3.149^{+0.041}_{-0.028}$. 
These values are incompatible with evolutionary models of single rotating stars, with the star rotating too rapidly for its evolutionary stage, and being undermassive for its luminosity. We conclude that $\theta$~Sco~A is most probably the product of a binary merger. 
\end{abstract}

\begin{keywords}
polarization -- techniques: polarimetric -- stars: evolution -- stars: rotation -- binaries: close
\end{keywords}



\section{Introduction}

Intrinsic stellar polarization was first predicted by \cite{chandrasekhar46}. He determined that, for a pure electron-scattering atmosphere, the radiation emerging from the stellar limb should be substantially linearly polarized. For a spherical star this polarization will average to zero, but a net polarization should result if there is a departure from spherical symmetry. \cite{Harrington} suggested that the distortion of a star due to rapid rotation could result in observable polarization, although calculations with more realistic non-grey stellar-atmosphere models, taking into account both scattering and absorption  \citep{Code, Collins, Sonneborn}, led to predictions of lower polarization levels at visible wavelengths. Consequently, it is only as a result of recent advances in instrumentation that the polarization due to rotational distortion has been detected, first in the B7\,V star Regulus \citep{cotton17} and subsequently in the A5\,IV star $\alpha$~Oph \citep{bailey20b}.

The polarization produced by rotational distortion is identifiable through its distinctive wavelength dependence. In hotter stars, such as Regulus \citep{cotton17}, the polarization shows a reversal of sign in the optical, being parallel to the star's rotation axis at red wavelengths but perpendicular in the  blue. In cooler stars, such as $\alpha$~Oph, the optical polarization is always perpendicular to the rotation axis, but falls from a relatively high value around $\sim$400~nm to near zero at wavelengths
of $\sim$800~nm and longer.

The polarization produced by scattering in  stellar atmospheres is predicted to be strongly dependent on gravity \citep[e.g., Figure~2 and Supplementary Figure~4 in][]{cotton17}. 
This is because the mass opacity due to scattering processes is independent of density, while the opacity due to most continuum absorption processes is roughly proportional to density \citep{kramers23}. A lower density (as found in lower-gravity stars) therefore results in an increased importance of scattering relative to absorption, and hence in higher levels of polarization. This applies to stellar polarization due to rotational distortion \citep{cotton17,bailey20b}, to photospheric `reflection' in binary systems \citep{bailey19a,cotton20}, and to non-radial pulsation \citep{cotton21}. In all these cases, giant stars should show higher polarization than main-sequence stars if other factors are equivalent. 

In this paper we test that prediction for the case of stellar rotation.
Our target, $\theta$~Scorpii (Sargas, HD~159532), was selected as a bright, rapid rotator ($V = 1.87$; \citealt{johnson66}),  classified as F1$\,$III by \citet{graygarr89}. \citet{vanbelle12} included it in a list of candidates for investigation with optical interferometry, and a detailed interferometric study of the star is reported by \citet{Souza}.

\section{Observations}

\label{sec_obs}

\begin{table*}
\caption{Summary of observing runs and telescope-polarization (TP) calibrations}
\label{tab:runs}
\tabcolsep 3.5 pt
\begin{tabular}{cl|ccrcccc|ccc|rr}
\toprule
\multicolumn{2}{c|}{} & \multicolumn{7}{c|}{Telescope and Instrument Set-Up$^a$}   &   \multicolumn{3}{c|}{Observations$^b$}   &   \multicolumn{2}{c}{Calibration$^{c}$}    \\
Run & \multicolumn{1}{c|}{Date Range$^d$} & Instr. &  Tel. & \multicolumn{1}{c}{f/} & Ap. & Mod. & Filt. & Det.$^e$ & $n$ & $\lambda_{\rm eff}$ &  Eff. & \multicolumn{1}{c}{$q_{\text{TP}}$} & \multicolumn{1}{c}{$u_{\rm TP}$} \\
 &  \multicolumn{1}{c|}{(UT)} &  &   &  & ($\arcsec$) &  &  &  &  & (nm) & ($\%$) & \multicolumn{1}{c}{(ppm)} & \multicolumn{1}{c}{(ppm)} \\
\midrule
2014MAY &  2014-05-12  &   HIPPI   & AAT  &   8  & 6.6 &  MT &  $g^{\prime}$& B   & 1 & 466.5 & 88.5 & $-$43.3 $\pm$ 0.9 & $-$53.2 $\pm$ 1.0 \\
2017AUG & 2017-08-12   &   HIPPI   & AAT  &   8  & 6.6 & BNS-E2 & $g^{\prime}$& B & 0 & & & $-$9.1 $\pm$	1.5 &	$-$2.6 $\pm$ 1.4\\
2018MAR & 2018-03-29 to 04-07   & HIPPI-2    & AAT  & 8*& 15.7 & BNS-E3 & 425SP & B   & 2 & 403.5 & 39.7 & 177.1 $\pm$ 2.8 & 25.1 $\pm$ 2.8 \\
& & & & & & & 500SP & B   & 2 & 441.5 & 70.5 & 142.6 $\pm$ 1.2 & 19.9 $\pm$ 1.2 \\
& & & & & & & $g^{\prime}$& B   & 1 & 467.4 & 82.6 & 130.1 $\pm$ 0.9 & 3.9 $\pm$ 0.9 \\
& & & & & & & $r^{\prime}$& R   & 1 & 633.4 & 81.1 & 113.3 $\pm$ 1.4 & 7.2 $\pm$ 1.4 \\
& & & & & & & 650LP & R   & 2 & 723.8 & 62.5 & 106.9 $\pm$ 1.9 & 10.4 $\pm$ 1.9 \\
2018JUL & 2018-07-25  & HIPPI-2  & AAT  & 8*& 11.9 & BNS-E4 & 425SP       & B   & 1 & 403.4 & 38.2 &  $-$5.6 $\pm$ 6.4 & 19.8 $\pm$ 6.3 \\
&                         &            &      &   &      &        & $V$           & B   & 1 & 534.5 & 95.4 & $-$20.3 $\pm$ 1.5 &  2.3 $\pm$ 1.5 \\
&                         &            &      &   &      &        & $r^{\prime}$& B   & 1 & 603.4 & 86.6 & $-$10.4 $\pm$ 2.2 &  3.7 $\pm$ 2.2 \\
2018AUG & 2018-08-18 & HIPPI-2   & AAT  & 8*& 11.9 & BNS-E5 & $V$           & B   & 1 & 535.5 & 95.2 & $-$20.3 $\pm$ 1.5 &  2.3 $\pm$ 1.5 \\

\bottomrule
\end{tabular}
\begin{flushleft}
Notes: \textbf{*} Indicates use of a 2$\times$ negative achromatic lens, effectively making the foci f/16. \textbf{$^a$} A full description, along with transmission curves for all the components and modulation characterisation of each modulator (`Mod.') in the specified performance era, can be found in \citet{bailey20a}. \textbf{$^b$} Mean values are given as representative of the observations made of $\theta$~Sco. Individual values are given in Table~\ref{tab:observations} for each observation; $n$ is the number of observations of $\theta$~Sco. \textbf{$^c$} The observations used to determine the TP and the high-polarization standards observed to calibrate position angle (PA), and the values of $\sigma_{\rm PA}$ are described in \citet{bailey15} (MAY2014), \citet{cotton19b} (AUG2017), and \citet{bailey20a} (other runs). \textbf{$^d$} Dates given are for observations of $\theta$~Sco and/or control stars. \textbf{$^e$} B, R indicate blue- and red-sensitive H10720-210 and H10720-20 photomultiplier-tube detectors, respectively.\\
\end{flushleft}
\label{tab:mod}
\end{table*}

\begin{table*}
\caption{Polarization observations of $\theta$~Sco (sorted by effective wavelength).}
\label{tab:observations}
\tabcolsep 4 pt
\begin{tabular}{llrrccrrrrrr}
\toprule
\multicolumn{1}{c}{Run}     & \multicolumn{1}{c}{UT}                  & Dwell & Exp. & Filt. & Det.$^a$ & $\lambda_{\rm eff}$ & Eff. & \multicolumn{1}{c}{$q$} & \multicolumn{1}{c}{$u$} & \multicolumn{1}{c}{$p$} & \multicolumn{1}{c}{$\theta$}\\
        &                       & \multicolumn{1}{c}{(s)} & \multicolumn{1}{c}{(s)} & & & \multicolumn{1}{c}{(nm)} & \multicolumn{1}{c}{(\%)} & \multicolumn{1}{c}{(ppm)} & \multicolumn{1}{c}{(ppm)} & \multicolumn{1}{c}{(ppm)} & \multicolumn{1}{c}{($^\circ$)}\\
\midrule
2018MAR & 2018-03-29 16:32:03 & 1683 & 1280 & 425SP & B & 403.5 & 39.8 &  $-$421.1 $\pm$ 13.3 &   32.5 $\pm$ 13.2 & 422.4 $\pm$ 13.3 &   87.8 $\pm$ \phantom{0}0.9\\
2018MAR & 2018-04-07 17:17:10 & 1736 & 1280 & 425SP & B & 403.3 & 39.6 &  $-$391.9 $\pm$ 13.2 & $-$31.4 $\pm$ 13.2 &   393.2 $\pm$ 13.2 &  92.3 $\pm$ \phantom{0}1.0\\
2018JUL & 2018-07-25 14:00:48 & 1064 & 640 & 425SP & B & 403.4 & 38.2 & $-$299.3 $\pm$ 16.5 &  $-$4.0 $\pm$ 16.4 &  299.3 $\pm$  16.5 &  90.4 $\pm$   \phantom{0}1.6\\
2018MAR & 2018-03-29 16:08:40 & 997 & 640 & 500SP & B & 441.8 & 70.7 & $-$242.0 $\pm$ \phantom{0}6.7 &  $-$0.7 $\pm$ \phantom{0}6.8 &   242.0 $\pm$   \phantom{0}6.8 &   89.9 $\pm$   \phantom{0}0.8\\
2018MAR & 2018-04-07 17:43:55 & 1185 & 730 & 500SP & B & 441.2 & 70.3 & $-$217.0 $\pm$ \phantom{0}6.6 & $-$4.9 $\pm$ \phantom{0}6.8  &  217.1 $\pm$   \phantom{0}6.7 &   90.7 $\pm$   \phantom{0}0.9\\
2014MAY & 2014-05-12 18:54:11 & 1257 & 720 & $g^{\prime}$ & B & 466.5 & 88.5 & $-$164.2 $\pm$ \phantom{0}3.2 &   12.4 $\pm$ \phantom{0}3.3 &  164.7 $\pm$   \phantom{0}3.2 &   87.8 $\pm$   \phantom{0}0.6 \\
2018MAR & 2018-03-29 15:51:38 & 986 & 640 & $g^{\prime}$ & B & 467.2 & 82.5 & $-$192.6 $\pm$ \phantom{0}2.8 &   22.9 $\pm$ \phantom{0}2.8 &  194.0 $\pm$   \phantom{0}2.8 &  86.6 $\pm$   \phantom{0}0.4\\
2018JUL & 2018-07-25 13:24:11 & 1072 & 640 & $V$ & B & 534.5 & 95.4 & $-$96.3 $\pm$ \phantom{0}4.1 &   17.3 $\pm$ \phantom{0}4.2 &   97.8 $\pm$   \phantom{0}4.2 &   84.9 $\pm$   \phantom{0}1.2\\
2018AUG & 2018-08-18 14:36:26 & 973 & 640 & $V$ & B & 534.9 & 95.2 &  $-$78.0 $\pm$ \phantom{0}4.0 &   25.2 $\pm$ \phantom{0}4.4 &   82.0 $\pm$   \phantom{0}4.2 &   81.0 $\pm$   \phantom{0}1.5\\
2018JUL & 2018-07-25 13:42:18 & 1045 & 640 & $r^{\prime}$ & B & 603.4 & 86.6 & $-$66.7 $\pm$ \phantom{0}7.0 &   17.1 $\pm$ \phantom{0}7.1 &  68.9 $\pm$   \phantom{0}7.0 &   82.8 $\pm$   \phantom{0}2.9\\
2018MAR & 2018-04-01 16:47:55 & 1001 & 640 & $r^{\prime}$ & R & 623.4 & 81.1 & $-$37.5 $\pm$ \phantom{0}3.8 &   33.4 $\pm$ \phantom{0}3.5 &   50.2 $\pm$   \phantom{0}3.6 &   69.2 $\pm$   \phantom{0}2.1\\
2018MAR & 2018-04-01 17:05:16 & 1012 & 640 & 650LP & R & 723.8 & 62.5 & $-$9.3 $\pm$ \phantom{0}5.5 &   35.5 $\pm$ \phantom{0}4.8  &   36.7 $\pm$   \phantom{0}5.2 &   52.3 $\pm$   \phantom{0}4.3\\
2018MAR & 2018-04-01 17:24:21 & 1162 & 800 & 650LP & R & 723.8 & 62.5 & $-$12.1 $\pm$ \phantom{0}7.4 &   51.7 $\pm$ \phantom{0}6.0 &   53.1 $\pm$   \phantom{0}6.7 &   51.6 $\pm$   \phantom{0}3.9\\
\bottomrule
\end{tabular}
\begin{flushleft}
{$^a$} B, R indicate blue- and red-sensitive H10720-210 and H10720-20 photomultiplier-tube detectors, respectively.
\end{flushleft}
\end{table*}

\begin{table*}
\caption{Observations of interstellar-control stars}
\label{tab:controls}
\tabcolsep 3.5 pt
\begin{tabular}{rlllrrrrrrrr}
\toprule
\multicolumn{1}{c}{Control}      & SpT   &   \multicolumn{1}{c}{Run}     & \multicolumn{1}{c}{UT}                  & Dwell & Exp. & \multicolumn{1}{c}{$\lambda_{\rm eff}$} & Eff. & \multicolumn{1}{c}{$q$} & \multicolumn{1}{c}{$u$} & \multicolumn{1}{c}{$p$} & \multicolumn{1}{c}{$\theta$}\\
\multicolumn{1}{c}{HD} &        &            &           & \multicolumn{1}{c}{(s)} & \multicolumn{1}{c}{(s)} & \multicolumn{1}{c}{(nm)} & \multicolumn{1}{c}{(\%)} & \multicolumn{1}{c}{(ppm)} & \multicolumn{1}{c}{(ppm)} & \multicolumn{1}{c}{(ppm)} & \multicolumn{1}{c}{($^\circ$)}\\
\midrule
HD~131342 & K2$\,$III   & 2018MAR & 2018-03026 14:22:44 & 1716 & 1280 & 475.8 & 86.1 & 17.9 $\pm$ \phantom{0}6.6 & 29.9 $\pm$ \phantom{0}6.7 & 34.8 $\pm$ \phantom{0}6.6 & 29.6 $\pm$ \phantom{0}5.5 \\
HD~138538 & K1/2$\,$III & 2018AUG & 2018-09-01 09:00:57 &  961  & 640 & 475.6 & 65.0 & 14.9 $\pm$ 12.2 & $-$20.4 $\pm$ \phantom{0}8.6 & 25.2 $\pm$ 10.4 & 153.0 $\pm$ 13.5 \\
HD~147584 & F9$\,$V     & 2018AUG & 2018-08-20 13:35:24 &  967  & 640 & 471.9 & 74.8 & 30.5 $\pm$ \phantom{0}7.9 & $-$13.3 $\pm$ \phantom{0}9.1 & 33.2 $\pm$ \phantom{0}8.5 & 168.2 $\pm$ \phantom{0}7.5\\
HD~160928 & A3$\,$IVn   & 2018MAR & 2018-03-29 17:03:49 & 1721 & 1280 & 462.9 & 80.7 & 36.0 $\pm$ \phantom{0}7.6 & $-$17.2 $\pm$ \phantom{0}7.9 & 39.9 $\pm$ \phantom{0}7.7 & 167.2 $\pm$ \phantom{0}5.6\\
                &   & 2018MAR & 2018-03-29 17:32:31 & 1695 & 1280 & 462.8 & 80.6 & 23.3 $\pm$ \phantom{0}7.8 & $-$5.6 $\pm$ \phantom{0}7.5 & 23.9 $\pm$ \phantom{0}7.7 & 173.2 $\pm$ \phantom{0}9.7\\
HD~153580 & F5$\,$V     & 2018AUG & 2018-09-01 09:28:47 & 3852 & 1040 & 468.5 & 60.7 & 14.8 $\pm$ \phantom{0}9.7 & 30.8 $\pm$ 10.5 & 34.2 $\pm$ 10.2 & 32.2 $\pm$ \phantom{0}8.9\\
HD~166949 & G8$\,$III   & 2018MAR & 2018-03-30 16:19:18 & 1704 & 1280 & 474.6 & 85.6 & \phantom{0}4.4 $\pm$ 12.2 & 10.8 $\pm$ 12.3 & 11.7 $\pm$ 12.3 & 34.0 $\pm$ 32.8\\
                &   & 2018MAR & 2018-03-30 16:47:44 & 1709 & 1280 & 474.4 & 85.6 & \phantom{0}$-$0.7 $\pm$ 12.5 & $-$17.7 $\pm$ 11.7 & 17.8 $\pm$ 12.1 & 133.9 $\pm$ 24.1\\
HD~169586 & G0$\,$V     & 2017AUG & 2017-08-12 15:01:10 & 4252 & 3200 & 473.1 & 89.3 & 155.7 $\pm$ 11.0 & \phantom{0}5.0 $\pm$ 10.5 & 155.8 $\pm$ 10.8 & \phantom{00}0.9 $\pm$ \phantom{0}1.9\\
HD~174309 & F5      & 2018MAR & 2018-03-30 18:33:06 & 1715 & 1280 & 467.1 & 82.5 & $-$138.2 $\pm$ \phantom{0}9.4 & $-$200.3 $\pm$ \phantom{0}9.6 & 243.3 $\pm$ \phantom{0}9.5 & 117.7 $\pm$ \phantom{0}1.1\\
HD~182369 & A4$\,$IV    & 2018MAR & 2018-03-23 18:29:51 & 1108 &  640 & 463.6 & 81.0 & $-$120.3 $\pm$ \phantom{0}7.5 & $-$9.4 $\pm$ \phantom{0}7.1 & 120.6 $\pm$ \phantom{0}7.3 & \phantom{0}92.2 $\pm$ \phantom{0}1.7\\
\bottomrule
\end{tabular}
\begin{flushleft}
Notes: All control-star observations were made with the SDSS g$^\prime$ filter and B photomultiplier tube. The same aperture as used for the $\theta$~Sco observations in the same run was used. Spectral types (SpT) are from SIMBAD.\\
\end{flushleft}
\end{table*}

Between March 2014 and August 2018 we obtained 13 high-precision polarimetric observations of $\theta$~Sco in seven photometric passbands, using HIPPI (the HIgh Precision Polarimetric Instrument; \citealp{bailey15}) and its successor HIPPI-2 \citep{bailey20a} on the 3.9-m Anglo-Australian Telescope (AAT) at Siding Spring Observatory. Standard operating procedures for these instruments were followed, with the data reduced using procedures described (for HIPPI-2) by \citet{bailey20a}. 

Originally developed for sensitive exoplanet work of the kind most recently reported by \citet{bailey21}, both instruments used Ferro-electric Liquid Crystal (FLC) modulators to achieve the rapid, 500~Hz, modulation required for high precision. Observations were made using either blue-sensitive Hamamatsu H10720-210 or red-sensitive Hamamatsu H10720-20 photo-multiplier tubes (PMTs) as the detectors. 

For the HIPPI-2 observations our standard filter set, described in \citet{bailey20a}, was used; briefly, this includes 425- and \mbox{500-nm} short-pass filters (425SP, 500SP), SDSS $g^{\prime}$, Johnson $V$, SDSS $r^{\prime}$ and a 650~nm long-pass (650LP) filter. The one HIPPI observation used an Omega Optics version of the SDSS $g^{\prime}$ filter instead of the Astrodon versions used with HIPPI-2. The blue-sensitive PMTs were paired with most of the filters, with the red-sensitive PMTs used for the 650LP observations, and for one of the $r^{\prime}$ observations. 

The small amount of polarization arising in the telescope optics results in zero-point offsets in our observations. These are corrected for by reference to the straight mean of several observations of low-polarization standard stars, 
details of which are given in \citet{bailey20a}. Similarly, the position angle ($\theta$, measured eastward from celestial north) is calibrated by reference to literature measurements of high-polarization standards, also given in \citet{bailey20a}. A summary of the calibrations and each observing run is given in Table~\ref{tab:runs}.

\begin{figure}
    \centering
    \includegraphics[width=8.5cm]{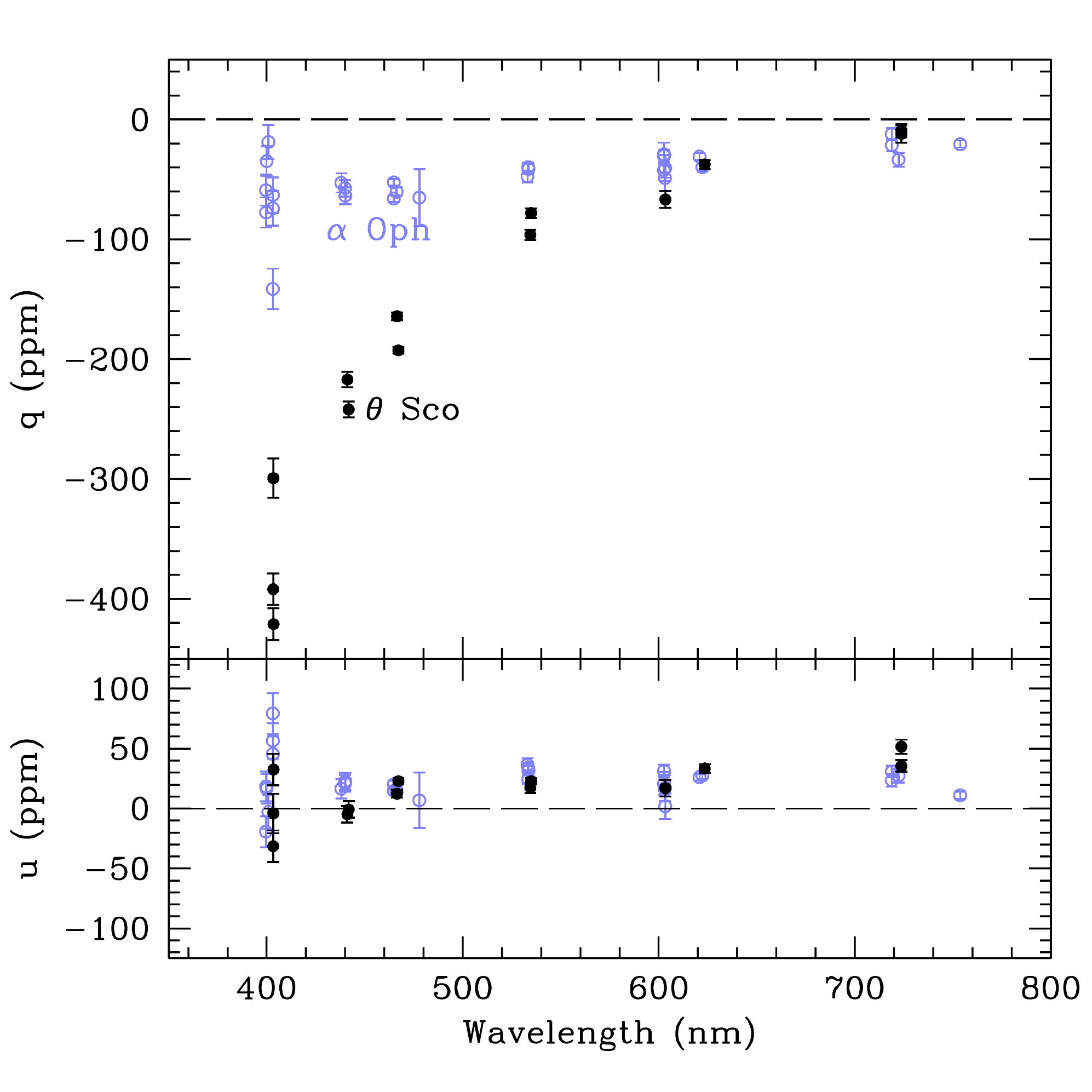}
    \caption{Polarization observations of $\theta$~Sco (black dots). The strong wavelength dependence seen in the $q$ Stokes parameter is of the form expected for rapid rotation. For comparison we plot, as blue circles, the polarization data on the rapidly rotating main-sequence star $\alpha$ Oph \citep{bailey20b}. Much higher polarization levels are seen in $\theta$~Sco. The $\alpha$~Oph data have been rotated to put the polarization into the negative q parameter as it is in $\theta$~Sco.}
    \label{fig:tetsco_points}
\end{figure}

Our observations of $\theta$~Sco are summarized in Table~\ref{tab:observations}; here the instrumental positioning error, resulting from inhomogeneities across the face of the FLC,
is included in the reported uncertainties. The data are presented in order of effective wavelength, and are plotted in \autoref{fig:tetsco_points}. The strong wavelength dependence seen in the normalized Stokes parameter $q$ ($=Q/I$) has the form expected for polarization resulting from rapid rotation, and the effect is much larger than seen in the rapidly rotating main-sequence star $\alpha$~Oph \citep{bailey20b}.

While the wavelength dependence is clearly seen in Figure~\ref{fig:tetsco_points} it is also apparent that there is more scatter in the observations than expected from the measurement errors. The errors ($\sim$ 3 ppm at $g^{\prime}$ band) are close to the highest precision achievable with the instrument. Under these circumstances we begin to see effects arising from changes in the instrument configuration, and from the imperfect characterization of the low polarization standard stars. In a more extensive study of polarization variability using these instruments \citep{bailey21} we found evidence for zero point differences between different observing runs by amounts of up to $\sim$10 ppm. These effects likely contribute to the scatter seen in the data. Any polarization variability in $\theta$~Sco over the period of observation will also contribute to the scatter.

\subsection{Interstellar control stars} \label{ics_sec}

Dust in the interstellar medium (ISM) can have a siginificant effect on the observed polarization. To determine the intrinsic stellar polarization it is therefore necessary to subtract the interstellar polarization, as part of the polarimetric modelling. The methodology can be tested by examination of normal stars thought not to be intrinsically polarized \citep{cotton16a}. \citet{cotton17b} developed an analytic model of interstellar polarization of stars close to the Sun by using control stars within 35$^\circ$ of a target. 
Data for a number control stars suitable for use with $\theta$~Sco  are available in the literature \citep{cotton17b, piirola20, marshall20}, and we have added to these through observations of the additional targets listed in Table~\ref{tab:controls}.

\begin{table}
\caption{New astrometric results, discussed in section  \ref{sec:hippo1}.   Magnitude differences are \dHp\ (\textit{Hipparcos}) and
$\Delta{y/I}$ (HRCam).}
\centering
\begin{tabular}{ll D{,}{\,\pm\,}{5} D{,}{\,\pm\,}{5} D{,}{\,\pm\,}{5} D{,}{\,\pm\,}{5} D{,}{\,\pm\,}{5} }
	\toprule
	\multicolumn{2}{c}{Parameter} &\multicolumn{1}{c}{\textit{Hipparcos}} &\multicolumn{1}{c}{\text{HRCam}}\\
	\midrule
Epoch & yr (CE) \T & \multicolumn{1}{c}{1991.25} &\multicolumn{1}{c}{2021.6}\\	
$\theta_{\rm B}$&$^\circ      \T          $& 274.3,0.13     & 97.54,0.78 \\  
$\rho        $  &$\arcsec                 $& 0.538,0.003    & 0.2447,0.0025    \\
$\Delta{m}$\B  &$\textrm{mag}            $&  3.27,0.01     & \multicolumn{1}{c}{3.32/3.69}\\
\dtbydt      \T &$^\circ/\textrm{yr}      $& -3.385,0.139   \\
\drbydt         &$\arcsec/\textrm{yr}  \B $& -0.0257,0.0021 \\
$\pi          $ &mas     \B               &9.90,0.26        \\
   \bottomrule
\end{tabular}
\begin{flushleft}

\end{flushleft}
\label{tbl:hippo}
\end{table}

\section{The binary $\theta$~Sco}

\subsection{Binarity revisited}
\label{sec:hippo1}

\citet{See96} reported the discovery, from Lowell Observatory, of a visual companion to
$\theta$~Sco~A ``of the 13th magnitude'', at a distance of 6\farcs2 in PA~322$^\circ$.
This would have been a challenging observation (the star's altitude would've been less
than 12$^\circ$), and \citet{ayres18} highlighted a number of inconsistencies in the
inferred properties of such a companion.

As part of the extensive pre-launch programme, \citeauthor{See96}'s observations were
included in the \textit{Hipparcos} preparatory data compilations.  One of several
possible solutions to the subsequent \textit{Hipparcos} astrometry agrees quite well with
\citeauthor{See96}'s report, and consequently was adopted in the final catalogue \citep{vanLeeuwen07x}, thereby
appearing to confirm the historical measurements. (Although \citeauthor{See96}'s
B-component magnitude estimate was much fainter than the \textit{Hipparcos} value of
$\sim$5$^{\rm m}$ -- in fact, his companion should ostensibly have been too faint to be
detected with the satellite -- this could have been consistent with his propensity for
over\-estimating binary-system brightness differences, often by large amounts.)

The agreement now appears to have been no more than an unfortunate
coincidence; subsequent direct visual observations have verified neither
\citeauthor{See96}'s report, nor its apparent partial confirmation by \textit{Hipparcos}.
As early as \citeyear{innes27}, \citeauthor{innes27} included the `discovery' in a list
of rejected observations after he failed to confirm
\citeauthor{See96}'s result. Of six rejected systems on the relevant page of the catalogue, four are attributed to \citeauthor{See96}.  More recently, visual observers \citet{kerr06} concluded
that ``the companion indicated by the \textit{Hipparcos} data does not exist [since they
would easily have detected it].  Moreover, the companion reported by \citeauthor{See96}
has not been re-observed [by skilled observers using comparable equipment at better
sites] and its existence is also in doubt.''   New speckle-interferometry results, discussed below, also rule out a 6\farcs2 companion  with a magnitude difference $\Delta{I} \lesssim 7.5$.

Although we believe that \citeauthor{See96}'s `discovery' must now be dis\-regarded, the
\textit{Hipparcos} photometric scans do, nevertheless, unambiguously indicate a visual binary (and, in
effect, now represent the discovery of a companion).  
We have therefore re-examined the raw \textit{Hipparcos} scans to investigate alternative
solutions, using an S-type differential analysis (cf.\ section~4.1.3 of
\citealt{vanLeeuwen07x}).   

While this paper was in preparation our work prompted new speckle interferometry with HRCam at
SOAR  \citep{tokovinin18}.     
The results (in particular, $\Delta{m}$) favour the new \textit{Hipparcos} solution 
that is listed in Table~\ref{tbl:hippo}, together with the average of four HRCam observations obtained on two nights a month apart (Tokovinin, personal communication).    This new \textit{Hipparcos} solution is consistent with results of
the visual double-star observers, and the inferred parallax is significantly more precise than
the original `new reduction' result.  

\subsection{The companion}
\label{sec:hippo2}

With its sub-arcsecond separation, the secondary was certainly included in the aperture of all photopolarimetric observations considered in this paper.   
Its properties are therefore of interest, if only for its potential as a contaminant.

Assuming, as a starting point, that $\Delta{V} = \Delta{H_{\rm P}}$, then a system magnitude of $V=1.87$ \citep{johnson66} implies 
$V_{\rm A, B} = 1.92, 5.19$, and provisional
$M(V)$ values of $-3.12$, $+0.16$ 
[for parallax $\pi = 9.90$~mas 
and differential extinction $E(B-V) = 0\fm005$; section  \ref{sec:grid}].

The secondary's absolute magnitude corresponds, very roughly, to spectral classifications 
$\sim$B8$\;$V, A0$\;$IV, or A3$\;$III
(e.g., \citealt{gray09};  brighter luminosity classes are excluded). 
On that basis we can next apply a differential-colour correction to the \textit{Hipparcos} photometry, taking $(V-I) \simeq +0.55, +0.10$ for the primary and secondary (observed and inferred; \citealt{johnson66}, \citealt{ducati01}).  Using the $(V-H_{\rm P}), (V-I)$ calibration from \citet{bessell00}, we obtain
$\Delta{V} \simeq \Delta{H_{\rm P}} + 0\fm07 = 3.34$ ($V_{\rm A,B} = +1.92, +5.26$), and final $M(V)$ values of $-3.12\pm0.06, +0.22\pm0.06$, where the 1-$\sigma$ ranges, derived from a simple Monte-Carlo analysis, incorporate the formal errors on the parallax (which remains the major source of uncertainty) and on \dHp, an assumed uncertainty of 0.01 on $V$, and a uniform distribution in $E(B-V)$ of
0:0.01 (section~\ref{sec:grid}).

  There is no detectable flux shortwards of 1500\AA\ in spectra obtained with the \textit{International Ultraviolet Explorer,} IUE (section~\ref{sec:grid}; the secondary would certainly have been included in these observations), implying
$\teff \lesssim 8.5$kK
(spectral type $\sim$A3 or later), whence $R \gtrsim 4\rsun$ in order to match $M(V)$.
The secondary's properties therefore appear to be broadly consistent with an $\sim$A3 giant star.   The HRCam $\Delta{m}$ results in Table~\ref{tbl:hippo} are quantitatively consistent with
such a companion.

\subsection{The orbit}

  The pair's angular separation at the \textit{Hipparcos} epoch  corresponds to a projected centres-of-mass separation of 54~au ($\sim$11\,600\rsun;
  orbital period \mbox{$\sim180 (a/54\text{~au})^{3/2}/\sqrt{\Sigma{M/5\msun}}$~yr}, where $a$ is the semi-major axis and $\Sigma{M}$ is the sum of the masses). The observations are insufficient to further constrain the orbital parameters, but
  the system appears to be wide in the sense that the B component is unlikely to have affected the primary's evolution.

\section{Modelling} 

\subsection{Polarization}
\label{Model_sec}

Our approach to modelling the polarization of a rotating star follows that previously used 
for $\alpha$~Oph (as described in detail by \citealt{bailey20b}) and Regulus \citep{cotton17}. We use a Roche model, with the variation of temperature over the stellar surface following from the  gravity-darkening law of \cite{espinosa11}. 
A rectangular grid of pixels is overlaid on the projected geometry, for given axial inclination $i$ and rotation $\omega$ ($=\Omega/\Omega_{\rm c}$, where $\Omega_{\rm c}$ is the critical angular rotation rate at which the centrifugal force balances gravity at the equator). For each pixel the local stellar temperature, gravity, and surface-normal viewing angle are calculated, and used to determine the intensity and polarization, as a function of wavelength.  Summing all the pixels gives the total intensity and polarization as a function of wavelength. The model co-ordinate frame is aligned with the rotation axis of the star in the plane of the sky, so that all the integrated polarization remains in the $q$ Stokes parameter (with the $u$ polarization being essentially zero).

The local intensity and polarization are interpolated from \textsc{atlas9} solar-composition stellar-atmosphere models computed for the local effective temperatures and gravities at 46 colatitudes, from 0$\degr$ to 90$\degr$ at 2$\degr$ intervals.
For each model atmosphere, the specific intensity and polarization are computed using a version of the \textsc{synspec} spectral synthesis code \citep{hubeny85,hubeny12} modified to do a fully polarized radiative-transfer calculation, using the \textsc{vlidort} code of \citet{spurr06}. 

The final integrated polarization results are then corrected to allow for the small contribution of the companion to the intensity. The companion is assumed to have a temperature of 8000~K and to have no intrinsic polarization.

\subsection{Stellar parameters}
\label{sec:grid}

The predicted polarization depends on four main parameters: the rotation rate ($\omega$), a reference temperature and gravity (e.g., polar values $T_{\rm p}$, $g_{\rm p}$), and the axial inclination ($i$). We use additional observational information to provide relationships between these parameters, thereby reducing a four-dimensional space to a two-dimensional grid.  We
take the projected equatorial rotation speed \vsini\ and distance $d$ (which together constrain $\omega$), and
the spectral-energy distribution (which constrains the temperature).
Then, for any specified ($\omega, i$) pairing, we can determine the (temperature, gravity) values that are uniquely consistent with all constraints, as described in detail in our study of $\alpha$~Oph \citep{bailey20b}, 

For simplicity (and with no important loss of information or sensitivity), we use the ratio of UV to $V$-band fluxes to determine the temperature.
The UV fluxes are from archival IUE spectra (SWP~48384 and LWP~26146, the only low-resolution, large-aperture spectra available), from which we obtain an observed 120--301~nm integrated flux of $9.85 \times 10^{-8}$~erg~cm$^{-2}$~s$^{-1}$.   We apply two corrections:  first, extinction for $E(B-V) = 0\fm005$ (based on the magnitude of the interstellar polarization; section~\ref{sec:resdis}), with a
\citet{seaton79} extinction law.  Secondly, we subtract an estimated B-component flux, based on an 8.0kK,
$\log{g}=4.0$ model, from \citet{howarth11}, scaled to the secondary's (dereddened) $V$~magnitude.   The corrections are each smaller than 10\%\ of the observed flux, 
and act in opposite directions.   We explore the dependence of the final results on the exact values as part of our sensitivity analysis (section~\ref{sec:best_fit_models}).

We also use the $V$-band results obtained in section~\ref{sec:hippo2}, and the
\textit{Hipparcos} distance of $101.0$~pc
(section~\ref{sec:hippo1}; the star is too bright to be included in \textit{Gaia} data releases available at the time of writing).
Finally, we adopt $\vsini = 91.7$~km~s$^{-1}$ from \citet{Souza};   we verified the plausibility of this value (and were unable to improve on it) by comparing synthesized spectra to an archival UVES spectrum obtained with the VLT.

\begin{figure}
    \centering
    \includegraphics[width=\columnwidth]{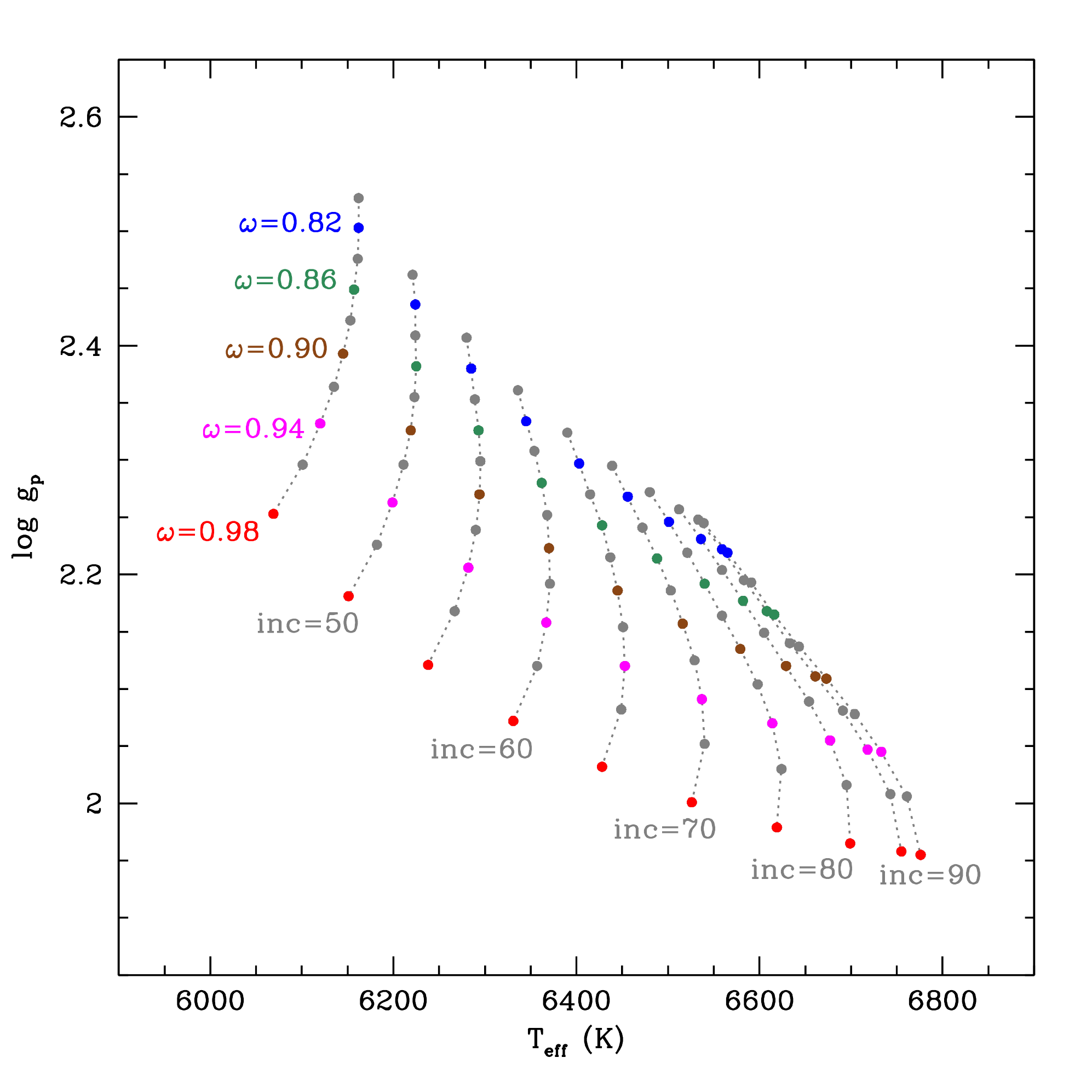}
    \caption{Temperature--gravity tracks for values of $\omega$ and $i$ for which the models discussed in section~\ref{sec:grid} reproduce the shape and level of the observed flux distribution, and $\vsini$.}
    \label{fig:grid}
\end{figure}

The resulting model grid for $\theta$~Sco~A is illustrated in Figure~\ref{fig:grid}. The grid covers $\omega$ from 0.8 to 0.99 in steps of 0.01 (only half these points are plotted on the figure), and inclinations from 45 to 90 degrees in 1-degree steps (only values at 5~degree steps are plotted on the figure). Each point in the grid represents a model star that reproduces the observed UV and $V$ flux levels (and the overall flux distribution) for the adopted $d$, $\vsini$ and the specified $i$, $\omega$. 

\section{Results}
\label{sec:resdis}

The polarization modelling described in section~\ref{Model_sec} provides predictions of the wavelength dependence of polarization for each of the models in the grid described in section~\ref{sec:grid}. We   compare these predictions to the observed polarization, corrected for interstellar polarization, to determine which model or models in the grid best match the data.

To make the comparison we integrate the model polarization over each filter bandpass \citep{bailey20a}.  Since the model polarizations are entirely in the $q$ Stokes parameter,  by aligning the observed and model polarization vectors we also determine the position angle of the star's rotation axis.   

\subsection{Interstellar polarization}

Interstellar polarization results from  dichroic scattering of light by aligned, non-spherical dust grains along the line of sight \citep{Davis}. It has a distinctive wavelength dependence, generally well approximated by the `Serkowski law':
\begin{equation}
    \frac{p(\lambda)}{p_{\rm{max}}} = e^{[-K\ln^2(\lambda_{\rm{max}}/\lambda)]}\label{eq:serk}
\end{equation}
\citep{Serkowski1971, Serkowski1973, serkowski75},
where $p(\lambda$) is the polarization at wavelength $\lambda$ and $p_{\rm{max}}$ is the maximum polarization,  occurring at  wavelength $\lambda_{\rm{max}}$. The normalizing constant $K$ has been found to be linearly related to $\lambda_{\rm{max}}$ \citep{wilking80}; \cite{whittet92} give
\begin{equation}
    K = (0.01 \pm 0.05) +(1.66 \pm 0.09) \lambda_{\rm{max}}
    \label{eq:kserk}
\end{equation}
\nopagebreak[4]
where $\lambda_{\rm{max}}$ is in $\mu$m. 

Because the wavelength dependence of the interstellar polarization is quite different to that of the rotational polarization it is possible to determine interstellar-polarization parameters in parallel with the fitting of stellar models to the observations. For each model in the grid we determined the difference between the modelled and observed polarization for each filter and fit a Serkowski curve, eqtn.~(\ref{eq:serk}), to these differences. 
The fits were carried out using the \textsc{curve\_fit} routine of the \textsc{python} package \textsc{scipy} \citep{scipy}. 

Although values for $\lambda_{\rm{max}}$ around 470~nm or less have been found for stars near to the Sun \citep{marshall16, cotton19b, bailey20b, marshall20}, $\theta$~Sco, at $d \simeq 100$~pc,  is near to the wall of the Local Hot Bubble, which is associated with a value of 550~nm \citep{cotton19b} -- this is also a typical value for the Galaxy as determined by \citet{serkowski75} and \citet{whittet92}. Consequently, we fixed $\lambda_{\rm{max}}$ to 550~nm, which in turn fixes $K$ (eqtn.~\ref{eq:kserk}). This leaves three fit parameters: $p_{\rm i}$ (equivalent to $p_{\rm max}$, but for fixed $\lambda_{\rm max}$), $\theta_{\rm i}$ (the position angle of the interstellar polarization), and $\theta_*$ (the position angle of the star's rotation axis).

\begin{figure}
\includegraphics[width=\columnwidth, trim={0.75cm 0cm 0.75cm 0cm},clip]{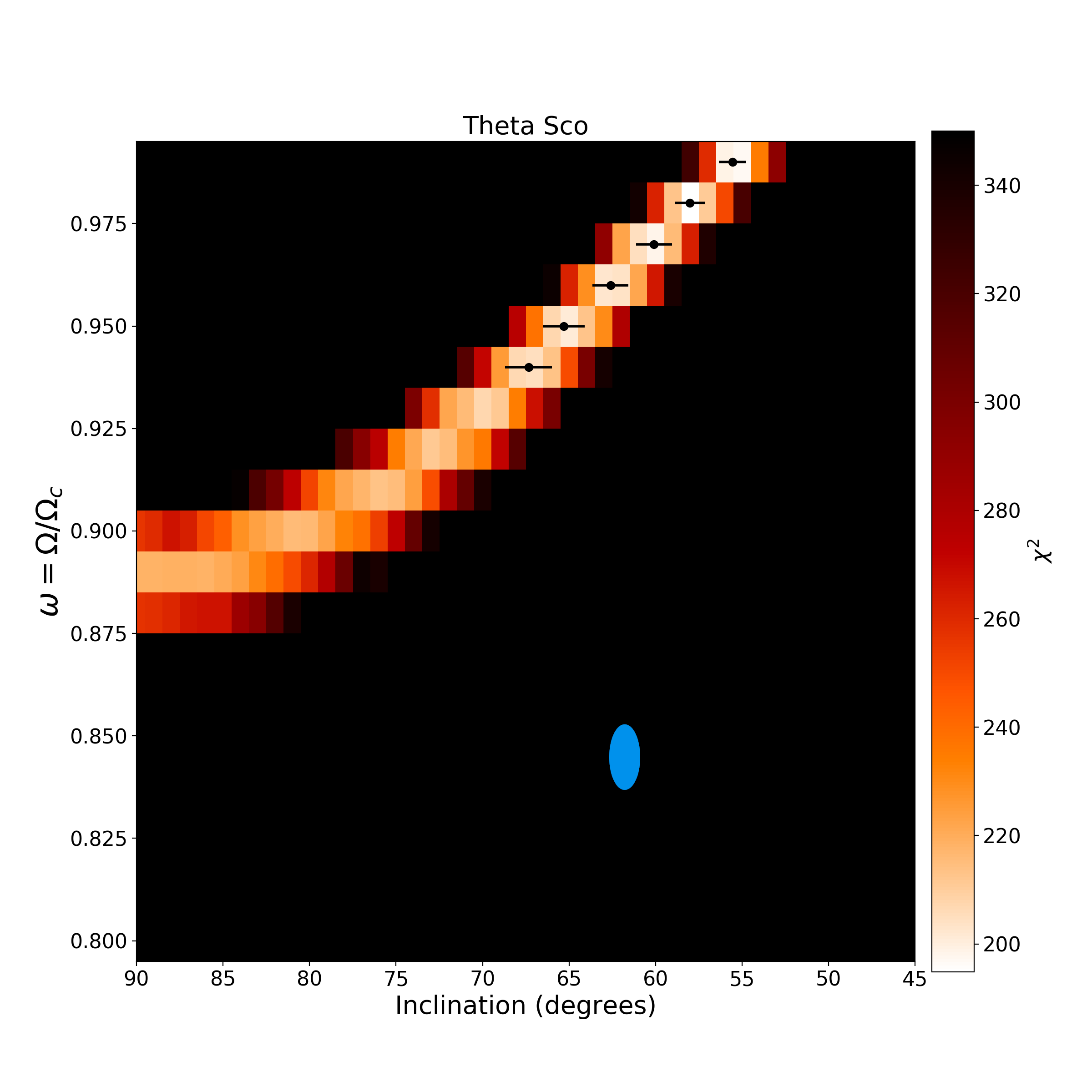}
\caption{Results of the $\chi^2$ analysis of the polarization of $\theta$~Sco.
The colour bar indicates the $\chi^2$ value (with lower $\chi^2$ indicating a closer fit of models and data). 
As described in section~\ref{sec:grid}, the parameter set for each cell in this grid represents a model star that matches the observed spectrum; the $\chi^2$ values show how well (or badly) each of these models also fits the observed polarization. Error bars on points in the best-fit `valley' are the result of bootstrap analyses for specific $\omega$ values (section~\ref{sec:best_fit_models}).  The blue dot indicates the solution obtained by \citeauthor{Souza} (\citeyear{Souza};  cf.\ section~\ref{sec_cps});   dot size  approximately matches the 1-$\sigma$ error bars they quote.}
 \label{fig:ts_chi}
\end{figure}

\subsection{Results: stellar parameters}
\label{sec:best_fit_models}

\begin{figure}
\includegraphics[width=\linewidth]{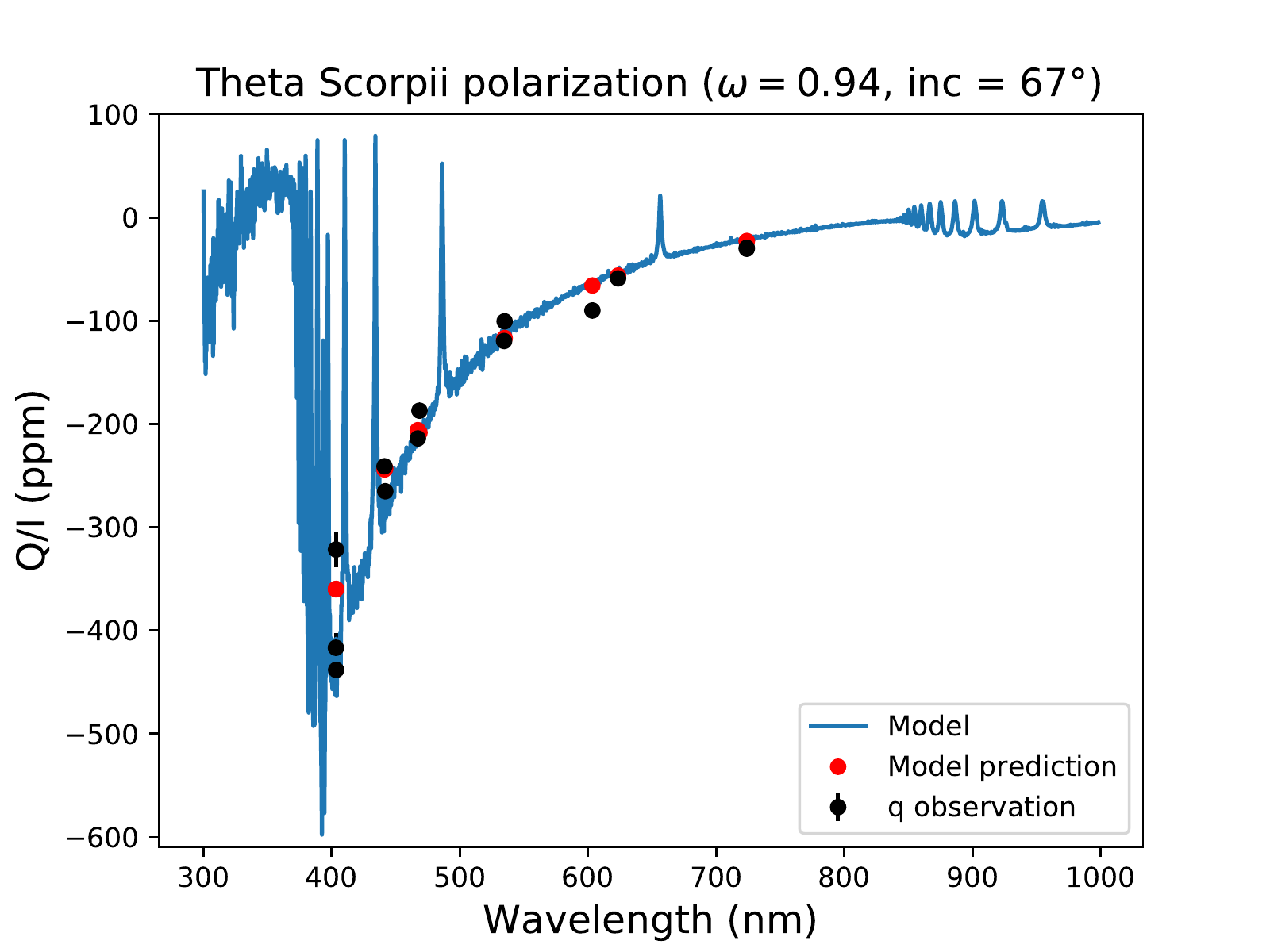}
\includegraphics[width=\linewidth]{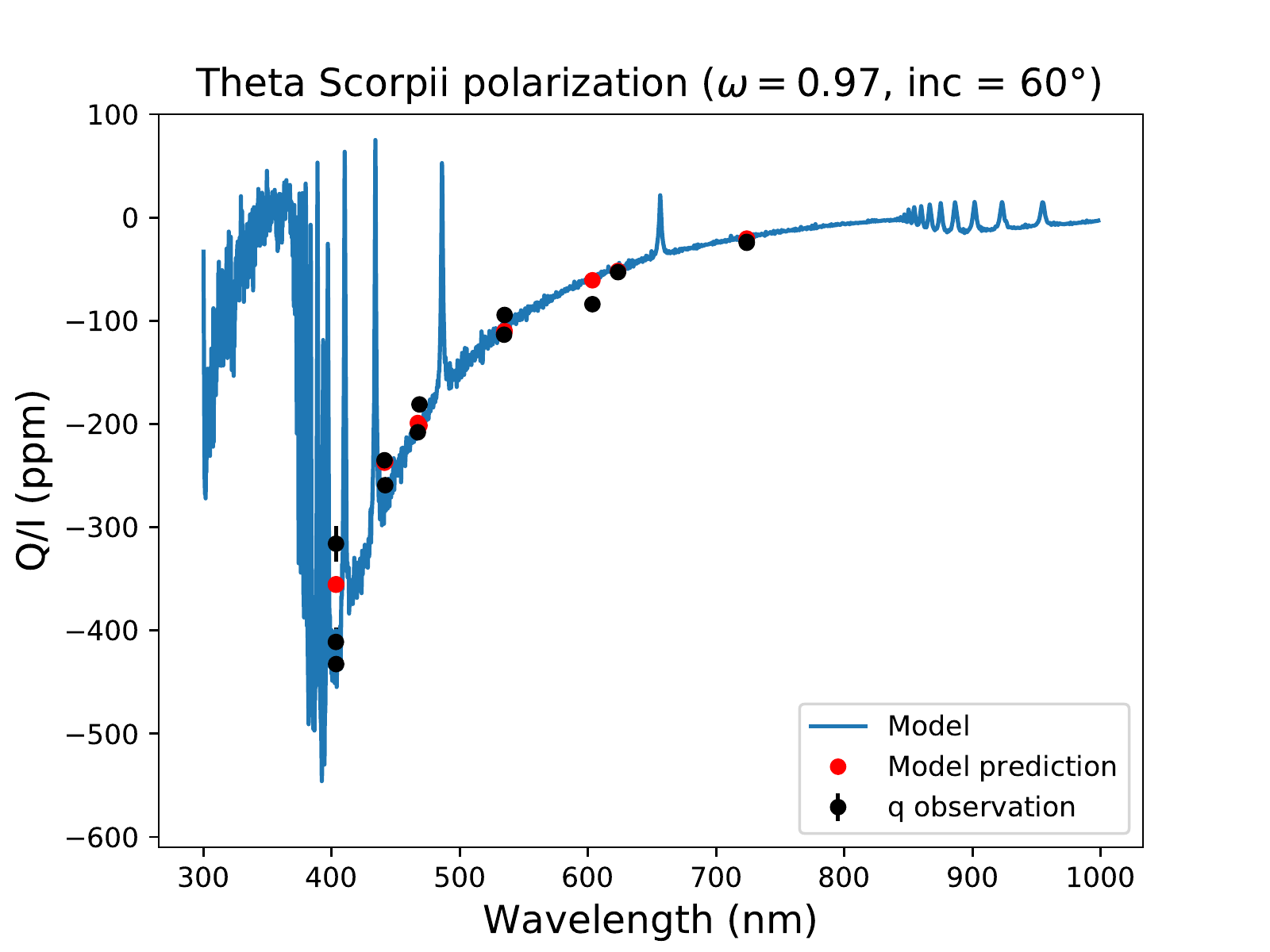}
\includegraphics[width=\linewidth]{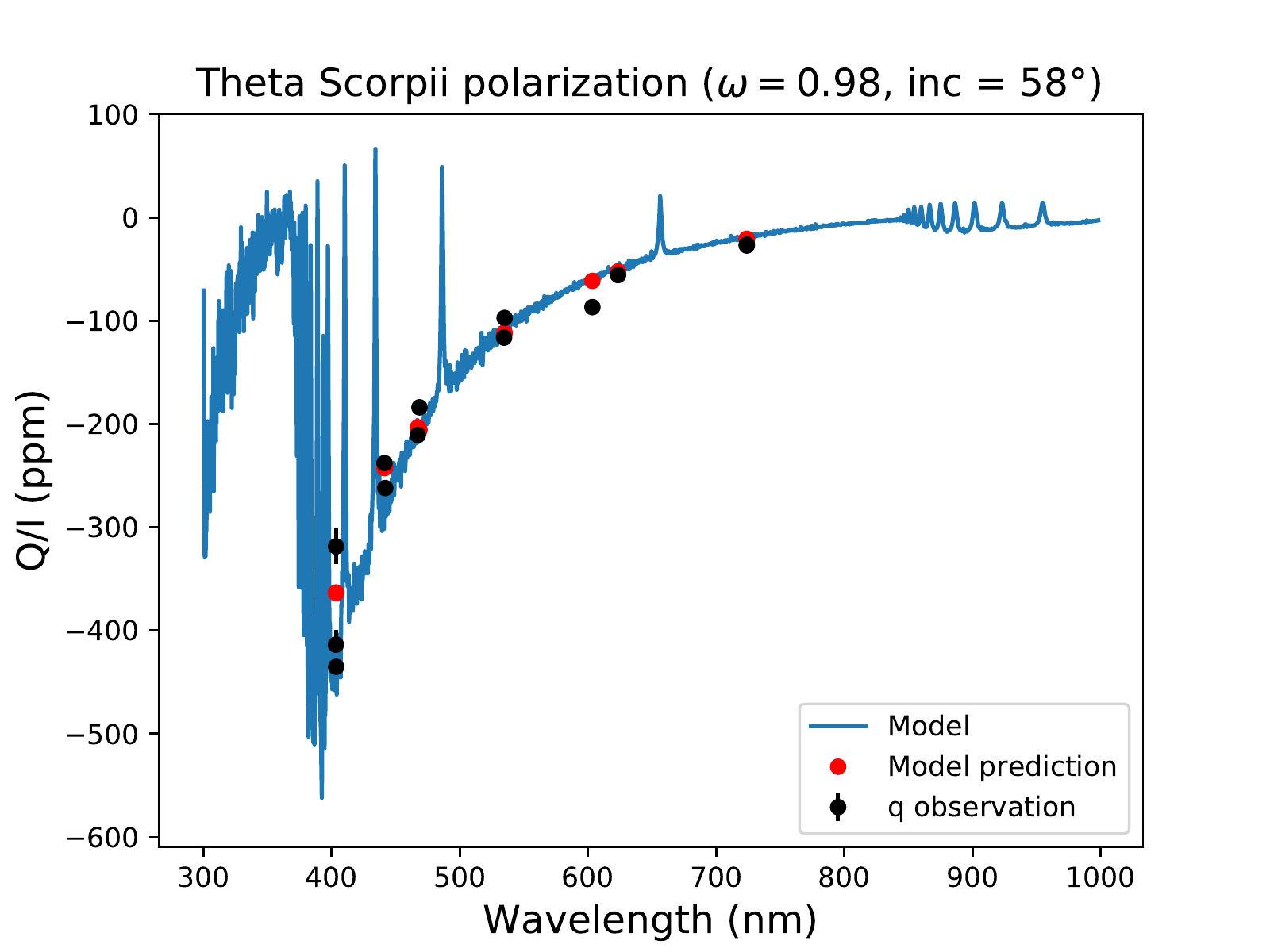}

\caption{Wavelength dependence of polarization for $\theta$~Sco observations, with best-fitting polarimetric model results for selected inclinations.  Black dots represent the observations, corrected for interstellar polarization and rotated;  polarimetric models are shown as
blue curves (full models) and red points (passband-integrated). Rotation of the observational data points into the model reference frame serves to essentially zero the $u$ Stokes parameter, leaving the polarization solely in the $q$ component.}
\label{fig:tS_fitting}
\end{figure}

Fit quality across the model grid was characterized by $\chi^2$, shown in
Figure~\ref{fig:ts_chi}. 
There is no single, well-defined $\chi^2$ minimum;  rather, a locus 
of low-$\chi^2$ values runs
across a rough diagonal in the $\omega/i$ plane, since these two parameters have similar effects on the polarization curve.  Figure~\ref{fig:tS_fitting} compares observed and modelled polarizations at several points along the $\chi^2$ `valley' to illustrate this near-redundancy.
The minimum-$\chi^2$ grid point is at $\omega = 0.98$, $i = 58^\circ$, but 
well-fitting models 
(within $1\sigma$ of the global minimum) can be identified at  all $\omega \gtrsim 0.94$.

\begin{figure}
\includegraphics[width=\linewidth]{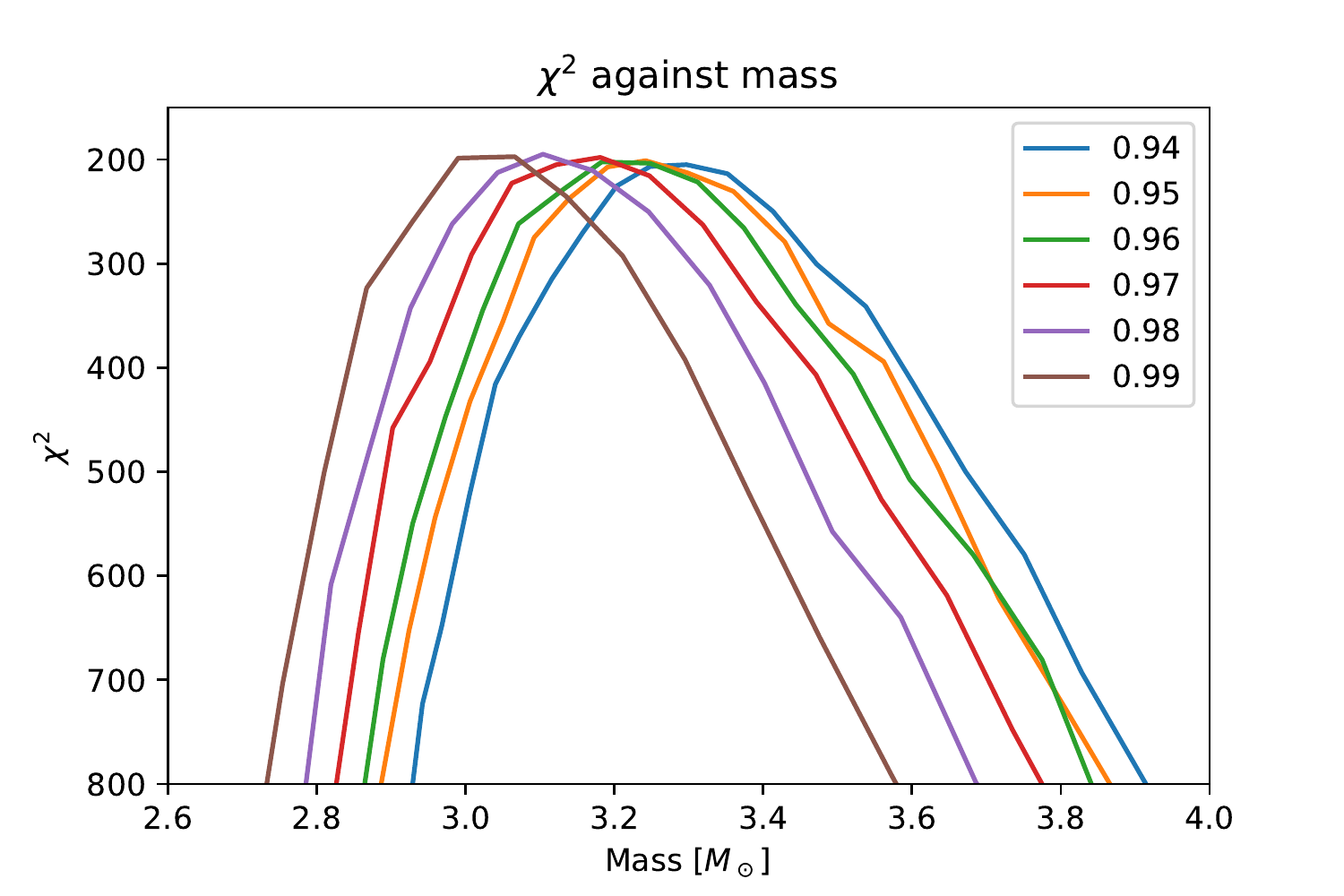}
\includegraphics[width=\linewidth]{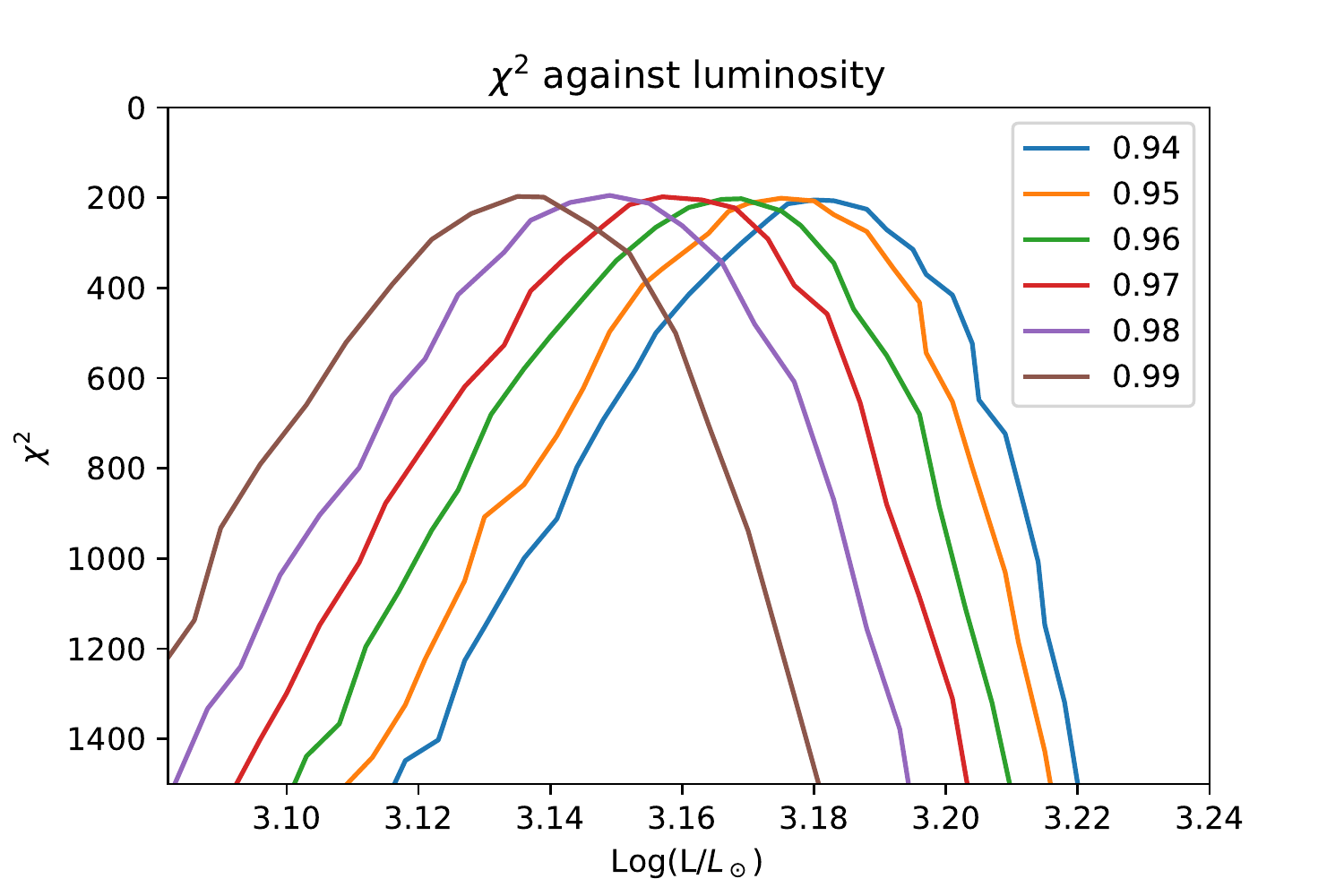}
\includegraphics[width=\linewidth]{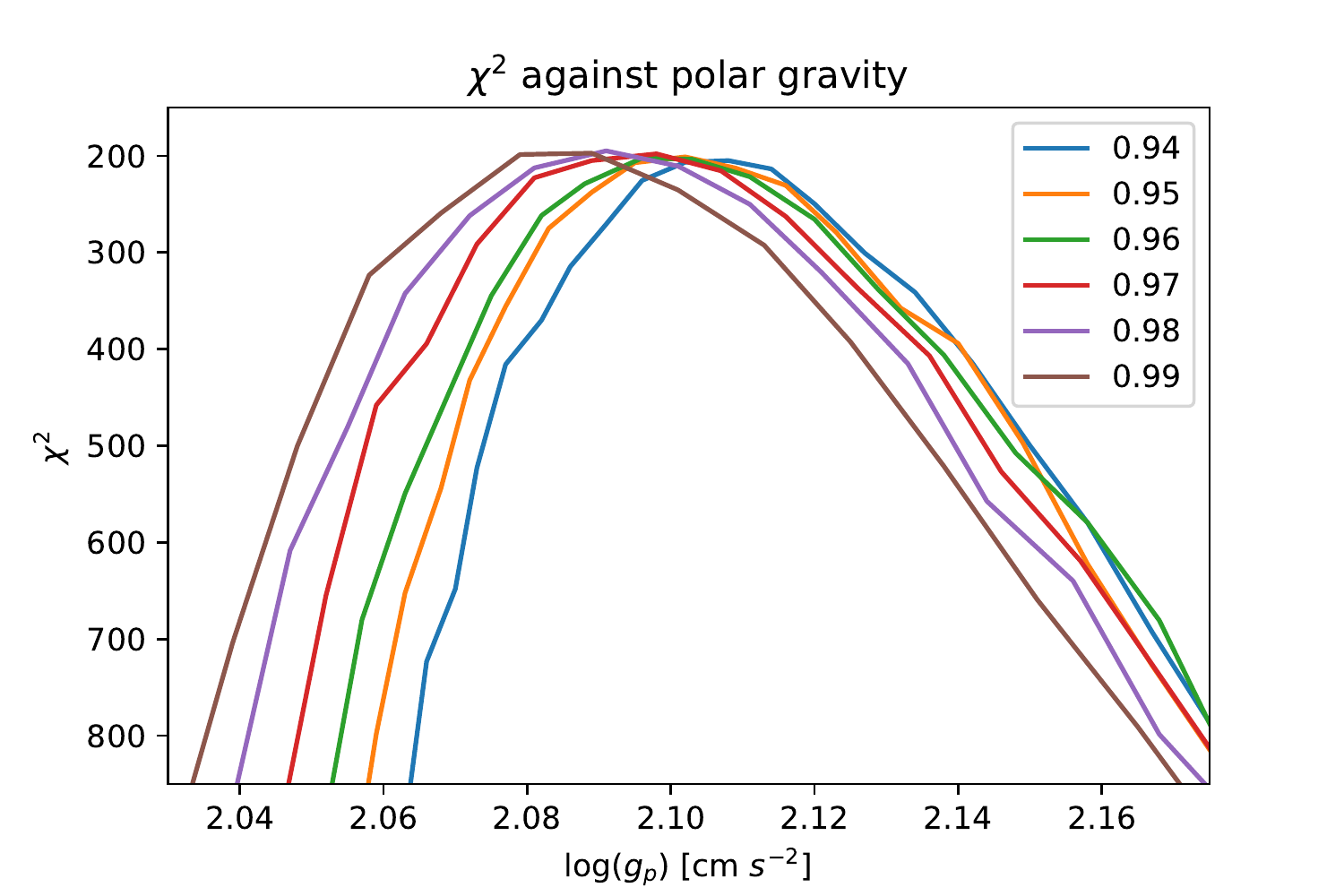}
\includegraphics[width=\linewidth]{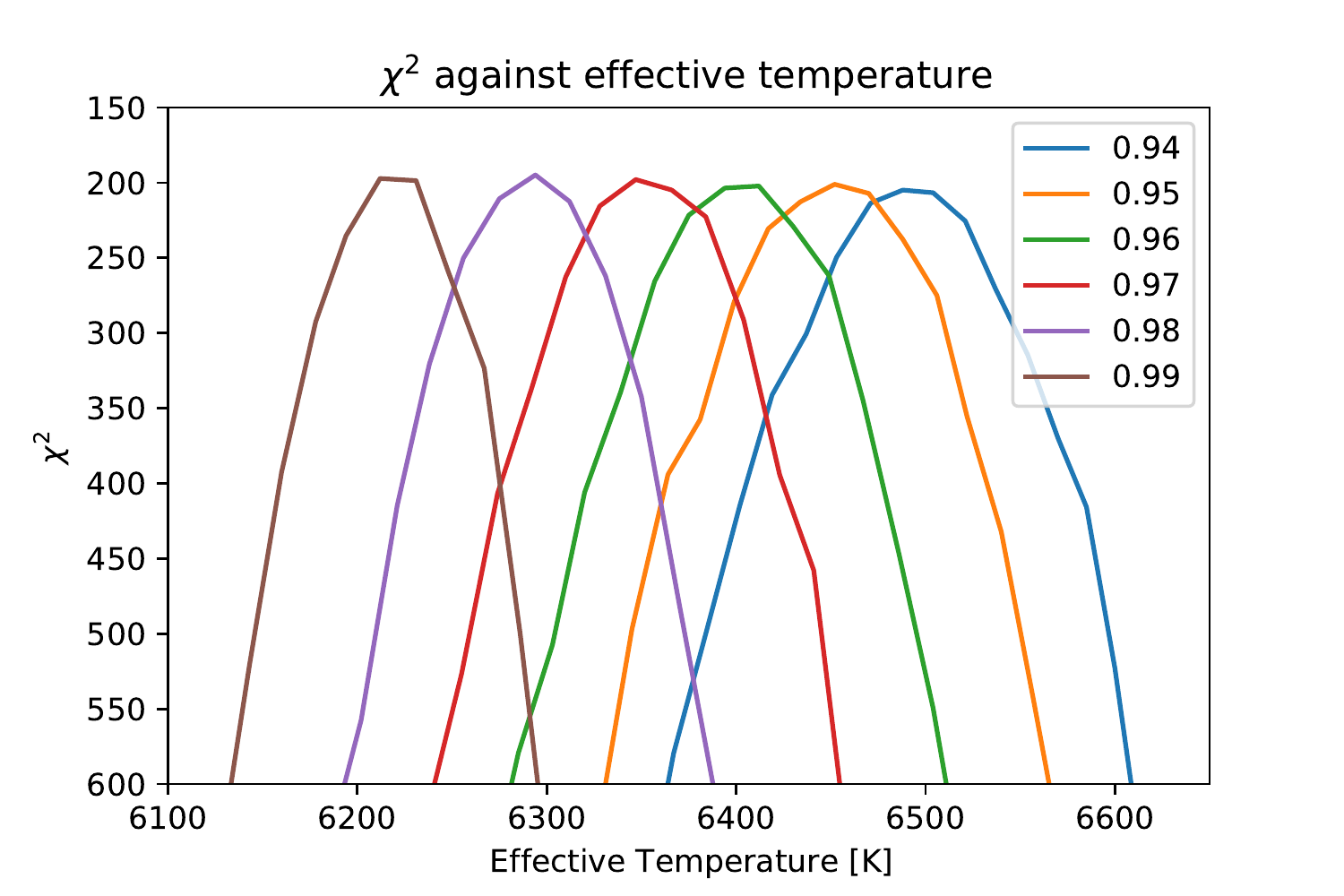}

\caption{Various stellar parameters plotted against $\chi^2$ values. Each plotted line represents a unique omega value as designated in the legend.}
\label{fig:tS_chi_parameters}
\end{figure}

Parameter uncertainties were estimated through bootstrapping (i.e., resampling with replacement; e.g., \citealt{Numerical_recipes}), for $\omega$ values
encompassing the 1-$\sigma$ range of best-fitting $\chi^2$ results (Table~\ref{tab:omega_inc}).
At each $\omega$,
a precise best-fit 
inclination was determined by spline interpolation in $\chi^2$, and its
1-$\sigma$ uncertainty estimated through 1000 bootstrapped replications. The resulting $\Delta\chi^2$ was then used to estimate errors on other parameters (cf.~Figure ~\ref{fig:tS_chi_parameters}). 
Results are summarized in Table~\ref{tbl:tS_parameters}. An advantage of this bootstrap procedure is that it bases the error determinations on the actual scatter in the data, rather than on the formal measurement errors that, as noted in section~\ref{sec_obs}, may be underestimated.

In addition to the statistical errors addressed by the foregoing procedures, systematic errors in parameter estimates will arise if the values adopted for observational constraints in section~\ref{sec:grid} are incorrect. We conducted simple sensitivity tests, the results of which are summarized in Table~\ref{tbl:sensit}.   As a baseline model we adopted the minimum-$\chi^2$ grid point 
($\omega=0.98$, $i=58^\circ$) and varied the parallax, UV flux, and $\vsini$ over reasonable ranges.   There are also smaller second-order effects resulting from displacements of the $\chi^2$ valley (Figure~\ref{fig:ts_chi}) with changing inputs. We estimated the additional errors due to these effects and added them in quadrature to the statistical errors from the bootstrap analysis to obtain the uncertainties listed in Table~\ref{tbl:tS_parameters}. 

One of the stellar parameters determined is the rotation period of the star, $P_{\text{rot}}$, for which we obtain 16.60 $^{+0.96}_{-1.03}$ days. $\theta$~Sco is not known to be a variable star. It is not listed in the General Catalogue of Variable Stars \citep{samus17}. We have examined the available space photometry from \textit{Hipparcos} and \textit{TESS} \citep[\textit{Transiting Exoplanet Survey Satellite,}][]{ricker15} and see no evidence for variability on the rotation period or any shorter period. $\theta$~Sco does not therefore show the periodic variability seen in some other rapidly rotating stars, in particular Be stars, and attributed to either non-radial pulsations \citep[e.g.][]{baade16} or rotational modulation \citep{balona20}.

\begin{table}
    \caption{Best-fit values of the inclination $i$ and their 1-$\sigma$ uncertainties, at selected  values of the rotation parameter, $\omega$}
    \label{tab:omega_inc}    \centering
    \begin{tabular}{cc}
         \toprule
          $\omega$ & $i$ (degrees)  \\
         \midrule
         0.99 & 55.54 $\pm$ 0.79 \\
         0.98 & 58.01 $\pm$ 0.87 \\
         0.97 & 60.09 $\pm$ 1.03 \\
         0.96 & 62.61 $\pm$ 1.03 \\
         0.95 & 65.31 $\pm$ 1.21 \\
         0.94 & 67.34 $\pm$ 1.35 \\
         \bottomrule
    \end{tabular}

\end{table}

\begin{table}
\caption{Best-fitting $\theta$~Sco~A parameters. The final column lists results from
\citet[their Table~2, $\omega$-model results]{Souza}, discussed in section~\ref{sec_cps}.}
\centering
\begin{tabular}{l  r@{\,}l c r@{\,}ll}
	\toprule
   Stellar parameter & \multicolumn{2}{c}{This Work}&\quad&\multicolumn{2}{c}{DdeS}\\
    \midrule 
    Inclination, $i$ [$^\circ$]& 58&$^{+9}_{-3}$&&61.8&$^{+0.8}_{-0.9}$\\ \\[-1em]
   $\omega$ & $\ge 0.94$& && $0.845$&$\pm 0.008$&[1]\\ \\[-1em]
    $T_{\text{eff}}$ [K]& 6294&$^{+224}_{-111}$&&6235&$^{+7}_{-8}$&[2]\\ \\[-1em]
    $\log(L/\lsun)$ & 3.149 &$^{+0.041}_{-0.028}$&&3.041&$\pm 0.006$&[1]\\ \\[-1em]
    $R_{\rm e}$ [\rsun]& 35.5&$^{+1.5}_{-2.3}$&&30.30&$^{+0.08}_{-0.09}$ \\ \\[-1em]
    $R_{\text{p}}$ [\rsun]&	26.3&$\pm$0.9 &&25.92&$\pm 0.09$&[1]\\ \\[-1em]
	Mass [\msun]&	3.10&$^{+0.37}_{-0.32}$ &&5.09&$^{+0.13}_{-0.14}$ \\\\[-1em]
    \loggp [dex cgs]& 2.091 &$^{+0.042}_{-0.039}$ &&2.317&$\pm 0.012$&[1]\\ \\[-1em]
    $P_{\text{rot}}$ [days] & 16.60 &  $^{+0.96}_{-1.03}$  && 14.74 &  $\pm$  0.16 &[1]\\ \\[-1em]
	$\theta_*$ [$^\circ$] & 3.3&$\pm1.0$ &&182.1&$^{+0.5}_{-0.4}$  \\
    \midrule 
    \multicolumn{2}{l}{Interstellar-polarization parameter} \\
	\midrule 
	$p_{\rm i}$ [ppm] & 43.7&$^{+6.2}_{-3.9}$ \\ \\[-1em]
	$\theta_{\text{i}}$ [$^\circ$] & 32.6&$^{+5.8}_{-5.2}$  \\
    \bottomrule
\end{tabular}
\begin{flushleft}
Notes: $\theta_*$  is the position angle of the stellar rotation axis, and $\theta_{\text{i}}$ is the position angle of interstellar polarization;  both have a $\pm180^\circ$ ambiguity in the case of the polarimetry (i.e. $\theta_*$ can be 183.3$^\circ$ or 3.3$^\circ$). \\
$[1]$ Uncertainty estimated by propagation of errors from data given by \citeauthor{Souza}, 
assuming 
$\sigma(R_{\rm p}) \simeq \sigma(R_{\rm e})$ 
where necessary.\\
$[2]$ \citeauthor{Souza} quote an ``average effective temperature'', presumably
$\int{\teff^{\ell}\text{d}A} \left/{\int{\text{d}A}}\right.$, 
where $\teff^\ell$ 
is the local effective temperature and d$A$ is an element of surface area.  
We have corrected this to match our definition, 
$\sigma_{\rm B}\teff^4 = 
L\left/{ \int{ \text{d}A } }\right. $
 \end{flushleft}
\label{tbl:tS_parameters}
\end{table}

\begin{table*}
\caption{Parameter sensitivity to fixed inputs, showing differences (model minus base) with respect to a reference model having
$\omega=0.98$, $i=58^\circ$, $\vsini=91.7$~km~s$^{-1}$.}
\centering
\begin{tabular}{l l r@{.}l cccccc}
\toprule
Parameter              &Unit       &     \multicolumn{2}{c}{Base}   &\multicolumn{2}{c}{Parallax} & \multicolumn{2}{c}{UV flux}&\multicolumn{2}{c}{\vsini}\\
&&\multicolumn{2}{c}{value}&$+1\sigma$&$-1\sigma$&$\times$0.95&$\times$1.05&$-5$ km/s&+5 km/s\\
\midrule
$R_{\rm p}/\rsun$       &           &     26&272 &$-0.658 $&$+0.614 $&$+0.614 $&$-0.599 $&$-0.223 $&$+0.129 $\\
$R_{\rm e}/\rsun$       &           &     35&475 &$-0.890 $&$+0.828 $&$+0.828 $&$-0.810 $&$-0.302 $&$+0.173 $\\
\loggp  & dex cgs   &     2&091  &$+0.011 $&$-0.010 $&$-0.010 $&$+0.010 $&$-0.045 $&$+0.044 $\\
\teff                   & kK        &     6&294  &$-0.004 $&$+0.005 $&$-0.063 $&$+0.062 $&$+0.019 $&$-0.018 $\\
$M/\msun$               &           &     3&104  &$-0.078 $&$+0.073 $&$+0.073 $&$-0.071 $&$-0.354 $&$+0.365 $\\
$\log{L/\lsun}$         &           &     3&149  &$-0.023 $&$+0.022 $&$+0.003 $&$-0.003 $&$-0.002 $&$+0.000 $\\
\bottomrule
\end{tabular}
\begin{flushleft}
Note: It may appear that, in principle, \teff\ should be independent of parallax;  in practice, small changes in the inferred gravity (and hence emergent model fluxes) result in changes of a few kelvin with distance. \\
\end{flushleft}
\label{tbl:sensit}
\end{table*}

\subsection{Results: interstellar polarization parameters}

\begin{figure*}
\includegraphics[clip, trim={3.3cm 0.2cm 2.8cm 1cm}, width=17.8cm]{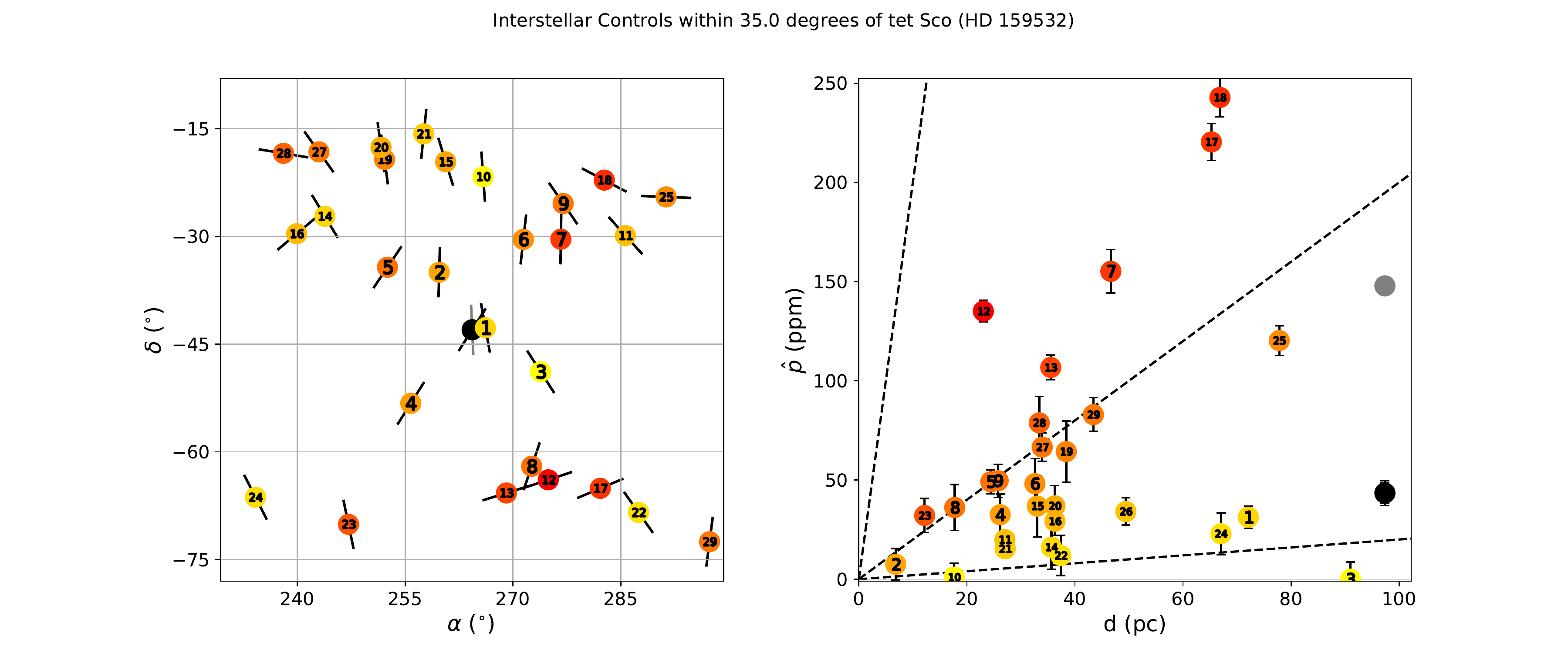}
\caption{Map (left) and $p$ vs. $d$ plot (right) of interstellar control stars within 35$^\circ$ of $\theta$~Sco. The control distances and co-ordinates were taken from SIMBAD.\\
Black pseudo-vectors on the data points indicate the position angles ($\theta_{\rm i}$), but not the magnitudes, of the interstellar polarizations. The effective wavelengths of the control observations have been used to standardise each to a wavelength of 450~nm (roughly corresponding to g$^\prime$), assuming a $\lambda_{max}$ of 470~nm -- which is appropriate since all are closer than $\theta$~Sco and thus probably not in the ``wall'' of the Local Hot Bubble \citep{cotton19b}.
The controls are colour coded in terms of $\hat{p}/d$ and numbered in order of their angular separation from $\theta$~Sco; they are: 1: HD~160928, 2: HD~156384, 3: HD~166949, 4: HD~153580, 5: HD~151680, 6: HD~165135, 7: HD~169586, 8: HD~165499, 9: HD~169916, 10: HD~160915, 11: HD~176687, 12: HD~167425, 13: HD~162521, 14: HD~146070, 15: HD~157172, 16: HD~143114, 17: HD~173168, 18: HD~174309, 19: HD~151504, 20: HD~151192, 21: HD~155125, 22: HD~177389, 23: HD~147584, 24: HD~138538, 25: HD~182369, 27: HD~131342, 28: HD~145518, 29: HD~147766, 30: HD~141937, 31: HD~186219. In the $p$ vs $d$ plot dashed lines corresponding to $\hat{p}/d$ values of 0.2, 2.0, and 20.0~ppm/pc are given as guides. The grey data-point is derived from the interstellar model in \citet{cotton17b} and the black data-point represents our best-fit interstellar values for $\theta$~Sco.}
 \label{fig:is_map}
\end{figure*}

The interstellar polarization parameters determined through the $\chi^2$ analysis can be checked through comparison with the interstellar polarizations of
control stars within 35$^\circ$ of $\theta$~Sco, illustrated in Figure~\ref{fig:is_map}. This region of the ISM is contained within the Local Hot Bubble, which extends $\sim$75--150~pc around the Sun and which has little dust, very patchily distributed \citep{bailey10, Frisch2010, cotton19b}.  The patchiness means that control-star observations cannot be used to  determine the interstellar values for $\theta$~Sco directly, but they can test whether the model  values (Table~\ref{tbl:tS_parameters}) are reasonable.  Figure~\ref{fig:is_map} shows that the polarization position angles of stars between 
declinations $-$30 and $-$60$^\circ$ are roughly aligned to $0 \pm 45^\circ$, consistent with the value found for $\theta$~Sco;  while
the magnitude of the inferred interstellar polarization for $\theta$~Sco is similar to  its closest neighbour on the sky, HD~160928.

Finally, we note that the magnitude of the rotationally-induced stellar polarization exceeds the interstellar component at all observed wavelengths, and dominates in the blue.

\begin{figure}
\includegraphics[width=\linewidth]{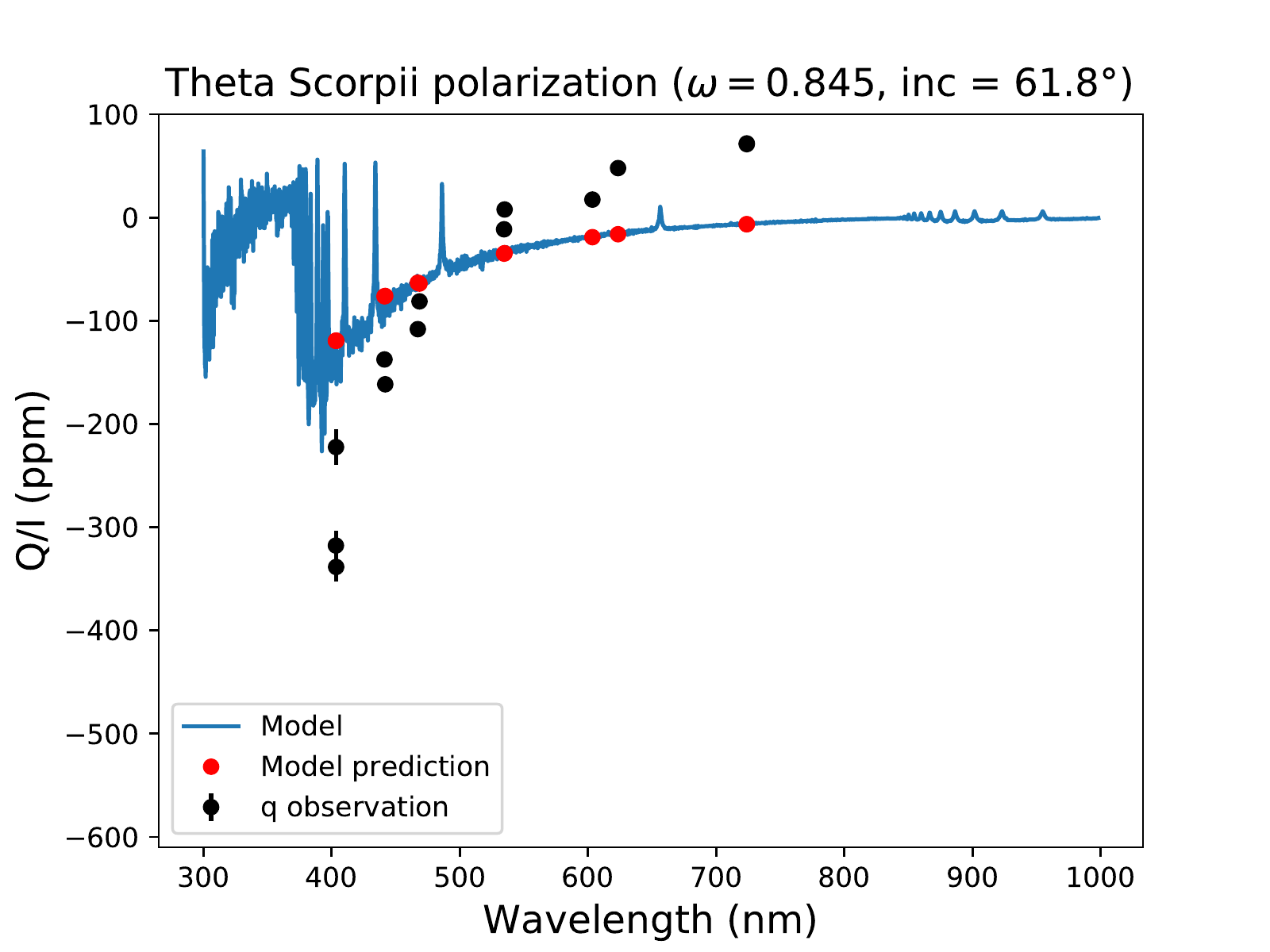}
\caption{Wavelength dependence of polarization for $\theta$~Sco observations (as Figure ~\ref{fig:tS_fitting}) compared with a polarization model based on the parameters of \citet{Souza}.    Black dots represent the observations, corrected for interstellar polarization and rotated;  polarimetric models are shown as
a blue curve (full models) and red points (passband-integrated).}
\label{fig:tS_fit_Souza}
\end{figure}

\subsection{Comparison with interferometry}
\label{sec_cps}

\citet{Souza} conducted  a detailed  study of $\theta$~Sco~A,
using optical interferometry supplemented with high-resolution spectroscopy.
A comparison of our results with theirs shows disappointingly poor agreement for  the inferred masses, and for other key stellar para\-meters
(Table~\ref{tbl:tS_parameters};  Figure~\ref{fig:ts_chi}).    The small difference in adopted distances does not account for the discrepancies (cf.\ Table~\ref{tbl:sensit});  nor are they attributable to contamination of the interferometric observations by the B~component. (Even if the then-unknown secondary was within the effective field of view of the interferometric instrumentation,  $\sim$0\farcs12--0\farcs15, at the time of observations (2016), it would not be expected to influence 
the interferometric results strongly; Domiciano de Souza, personal communication.)

We have been unable to identify any aspects of our analysis which are likely to account for the differences. 
In particular, the observed \mbox{polarization} cannot be reproduced without near-critical \mbox{rotation} 
(\mbox{$\omega \gtrsim 0.9$};  Figure ~\ref{fig:ts_chi}), which requires 
$M\sim 3\msun$. Figure ~\ref{fig:tS_fit_Souza} shows the intrinsic polarization predicted by our model for the \citeauthor{Souza} parameter set ($\omega = 0.845$;
we quote numerical results from the `$\omega$-model' solution given in Table~2 of \citet{Souza}, for consistency with our analysis).  Comparison with observations shows differences that are too large to be accommodated by stochastic errors, or by plausible uncertainties in polarization arising in either the interstellar medium or, potentially, the B~component.   

Although it is, of course, possible that our modelling code is in error, it has been tested against independent third-party calculations without giving any cause for concern \citep{cotton17}.
There are, however, indications that the posterior probability distributions generated by the \citeauthor{Souza} MCMC model-fitting procedure may not be fully reliable.   For example, a distance of
\mbox{$91.16\pm12.51$~pc} was used as a prior in their analysis;   since the interferometry gives, essentially, an angular-diameter measurement, we would  expect uncertainties on inferred radii of not less than
$12.51/91.16\simeq14\%$, yet the quoted error on $R_{\rm e}\text{ is actually}<0.3$\%.

The \citeauthor{Souza} $\teff$ determination, $6215^{+7}_{-8}$~K also appears to 
be remarkably precise for a measurement based on the synthesis of only a 26-nm stretch of rectified spectrum, and while they give no error on 
\vsini\ -- which, like \teff, must be constrained primarily by spectro\-scopy --
the upper limit given by $\sigma(v_{\rm e}), \simeq1.0$~km~s$^{-1}$, again seems unexpectedly small.

(\citet{Souza} separately list results based solely on
inter\-ferometry in their Table~3, which includes values for the equatorial rotation 
velocity with stated accuracies of $\sim$10\%;  it is unclear to us how this parameter can
be determined at all using only inter\-ferometry.
Other parameter values quoted in the Table as `not constrained' are 
indeed completely 
indeterminate from inter\-ferometry alone, so that the numerical values given there, and
their uncertainties, are arbitrary.)

Our own spectrum-synthesis calculations show that rectified spectra computed for $\pm{1}\sigma$ changes in \teff\ or $v_{\rm e}$ differ by \mbox{$\sim0.2$\%} in the core of H$\alpha$ (and by much less elsewhere), which is an order of magnitude smaller than the corresponding O$-$C (or likely rectification uncertainties), and is comparable to the purely statistical errors in the observed spectra.
Even as purely formal errors, we therefore suspect that the quoted \teff\ and $v_{\rm e}$ uncertainties may also be unrealistically small.

If measures of dispersion in at least some of the posterior distributions determined by \citet{Souza} are indeed too small, then measures of central tendency (i.e., parameter values) may also be open to question.  This is difficult to scrutinize directly since 
\citeauthor{Souza} delegated basic observables (such as angular diameters) to derived quantities, for which they did not propagate uncertainties.


\subsection{The potential utility of UV polarimetry}

As discussed in section~\ref{sec:best_fit_models}, $i$ and $\omega$ are not separately well-constrained for $\theta$~Sco (nor for $\alpha$~Oph, for which the situation is worse;  \citealt{bailey20b}), because these parameters have similar effects on the polarization curve over the wavelength range sampled. In this section we briefly examine prospects for resolving this redundancy through observations in the UV, such as allowed by the proposed \textit{Polstar} mission \citep{scowen21} -- which has been suggested for these types of observations \citep{jones21}.

To this end we ran a \textsc{synspec/vlidort} model for each 5$^\circ$ increment from Table~4 of
\citet{bailey20b} for $\alpha$~Oph, and from Table~\ref{tab:omega_inc} of the present paper for $\theta$~Sco, 
using appropriate $\omega$, $i$ values as listed there, with \teff\ and \loggp\ interpolated in the 
model grid in inclination for $\alpha$~Oph and in $\omega$ for $\theta$~Sco. The models were run over the 
wavelength range 100--300~nm, using the UV line lists `gfFUV' and `gfNUV' acquired from the \textsc{synspec} 
website and originally computed by \citet{kurucz99}.

%
%
%
\begin{figure*}
 \begin{minipage}{0.48\linewidth}
\centering
\includegraphics[clip,trim={0.9cm 0.0cm 0.4cm 1cm}, width=0.99\linewidth]{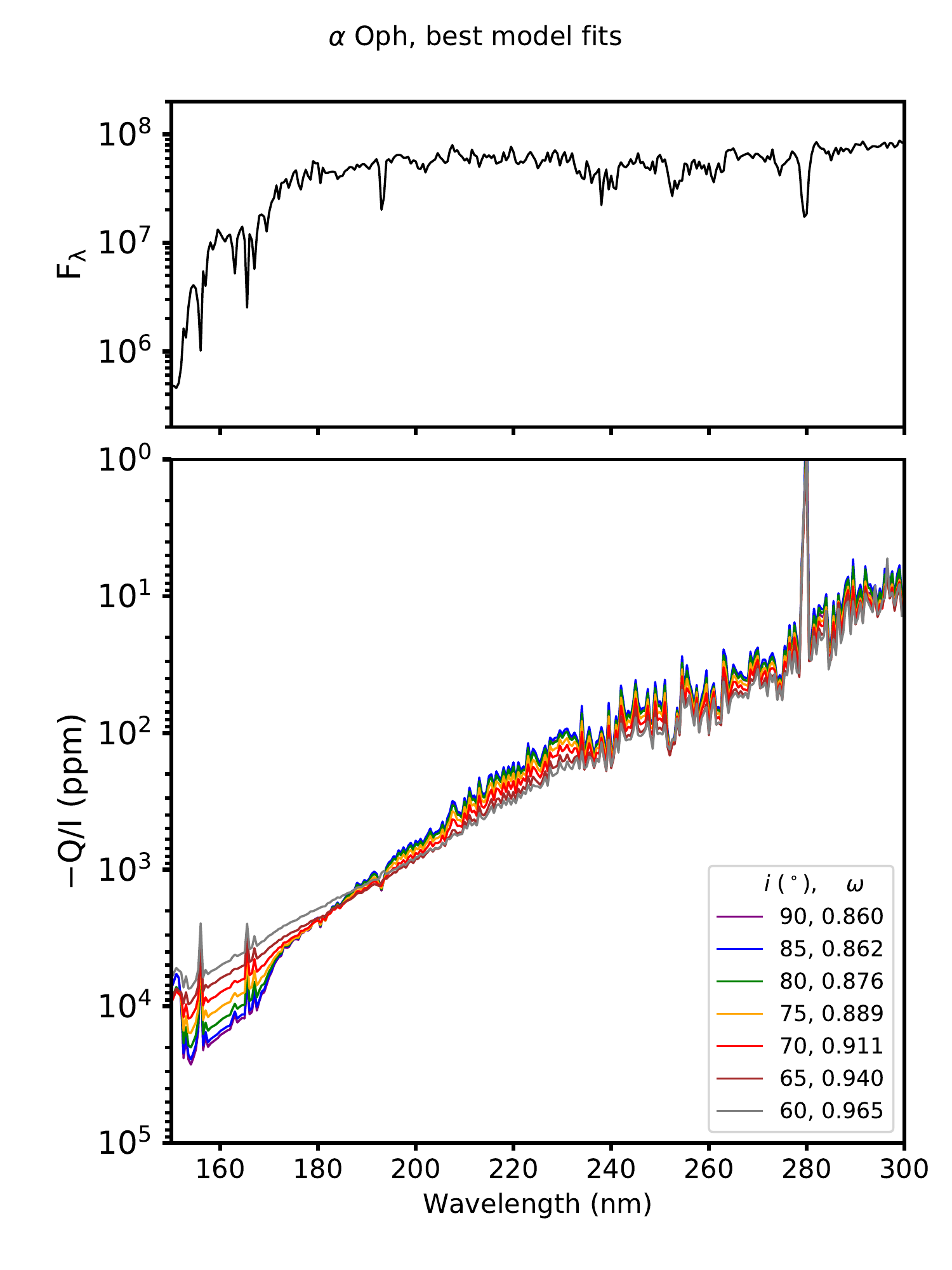}
\end{minipage}
\hfill
\begin{minipage}{0.48\linewidth}
\centering
\includegraphics[clip,trim={0.9cm 0.1cm 0.4cm 1cm}, width=\linewidth]{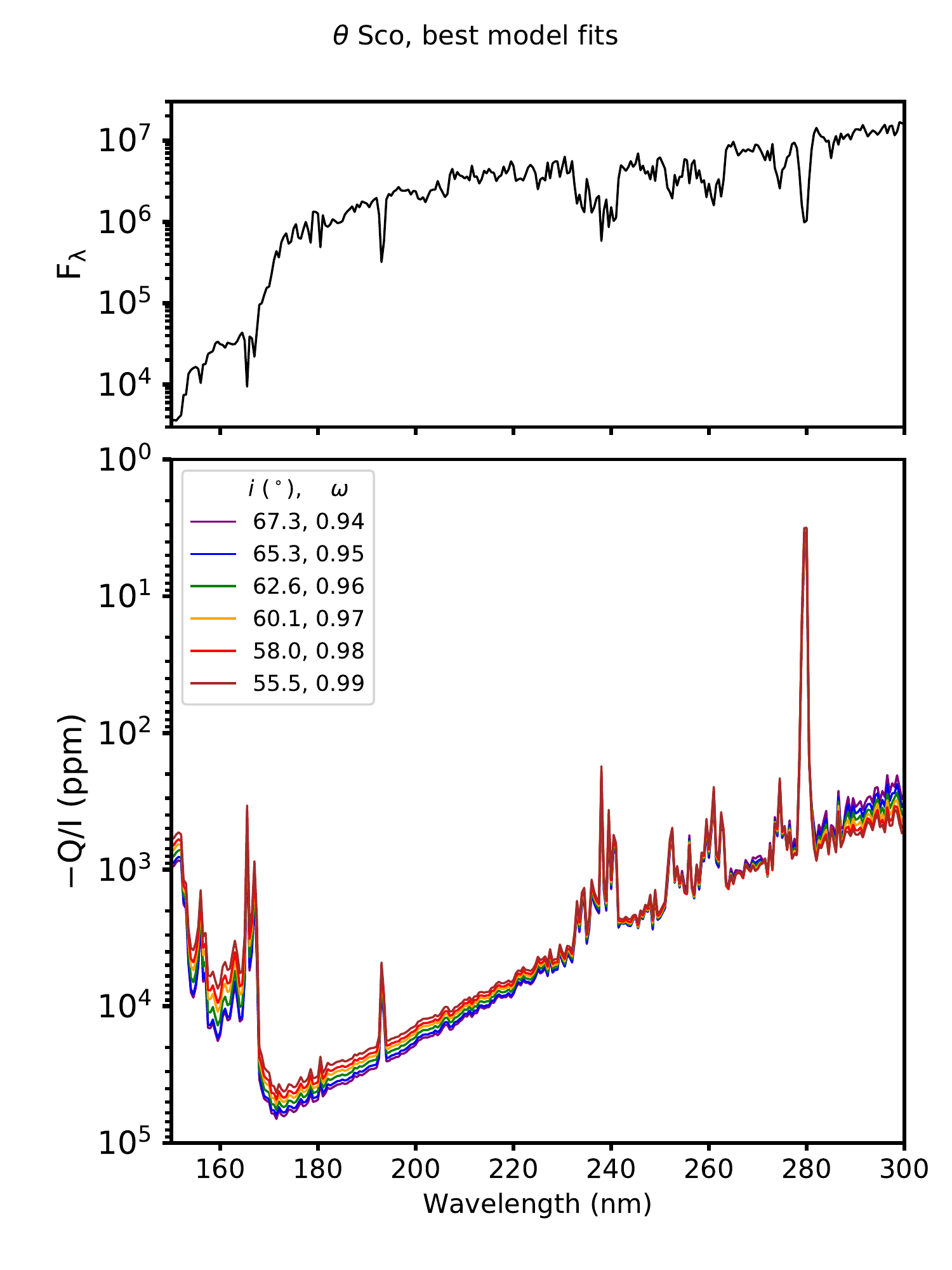}
 \end{minipage}
 \caption{Left-hand panels:  theoretical UV surface fluxes, $F_\lambda$, for $\alpha$~Oph (top, in units of erg~cm$^{-2}$~s$^{-1}$~\AA$^{-1}$), binned to 0.5~nm;  and polarization spectra (bottom), corresponding to the optical models given in Table~4 of \citeauthor{bailey20b} (\citeyear{bailey20b};  coloured lines). Note that the abscissa of both panels are on log scales, with the most negative $Q/I$ values at the bottom of the lower panel.\newline
 Right-hand panels: same, but for $\theta$~Sco.   Polarization spectra are shown for each of the models given in Table~\ref{tab:omega_inc} of the present paper.}
 \label{fig:stars_uv}
\end{figure*}

The results of the modelling are shown in 
Figure~\ref{fig:stars_uv}.
Below 300~nm $Q/I$ becomes increasingly more negative (the model $U/I$ is essentially zero; section~\ref{Model_sec}); 
this is similar to results for the highly inclined B-type stars modelled to 250~nm by \citet{Sonneborn}, and much further into the UV by \citet{collins91}. Polarization increases considerably with decreasing wavelength,  reaching a few per cent before the flux becomes negligible. 

For $\alpha$~Oph we see that, although the 85$^\circ$ and 90$^\circ$ models are nowhere easily separated, there are two regions that appear promising for distinguishing between the models at either end of the inclination range. In the longer-wavelength end of the range the low-inclination models are more polarized; between 200 and 240~nm the difference in polarization between the 60$^\circ$ and 90$^\circ$ models is roughly 100 ppm. Shortward of 180~nm, where the higher-inclination models are the more polarized, the difference between extreme inclinations is 1\%\ (10$^4$~ppm), though the very low flux here presents an impediment to utilising this region in practice for all but the brightest stars.

Models of $\theta$~Sco that are indistinguishable in the optical begin to diverge between 350 and 300~nm (Figure ~\ref{fig:tS_fitting}); this trend continues  down to the Mg\,II absorption lines at $\sim$280~nm 
(Figure ~\ref{fig:stars_uv}),
where the lowest-inclination model exhibits around 200~ppm more polarization than the highest-inclination model, making this a promising region for study. Some small (10--20~ppm) divergence in this region is also seen for $\alpha$~Oph. However, it is much stronger in $\theta$~Sco, in part because it has lower gravity and is more polarized in general. 

Between 230 and 280~nm the various $\theta$~Sco models are indistinguishable. Shortward of this range the higher-inclination models exhibit more polarization, such that below 190~nm the models are again distinct across the entire inclination range, at first differing by a fraction of a per~cent, rising to a few per cent at 170~nm; however, as an F-type star, the flux in this region is at best $\sim$5\%\ of what it is at 300~nm.

\section{Evolutionary status}
\begin{figure}
\includegraphics[width=8.25cm]{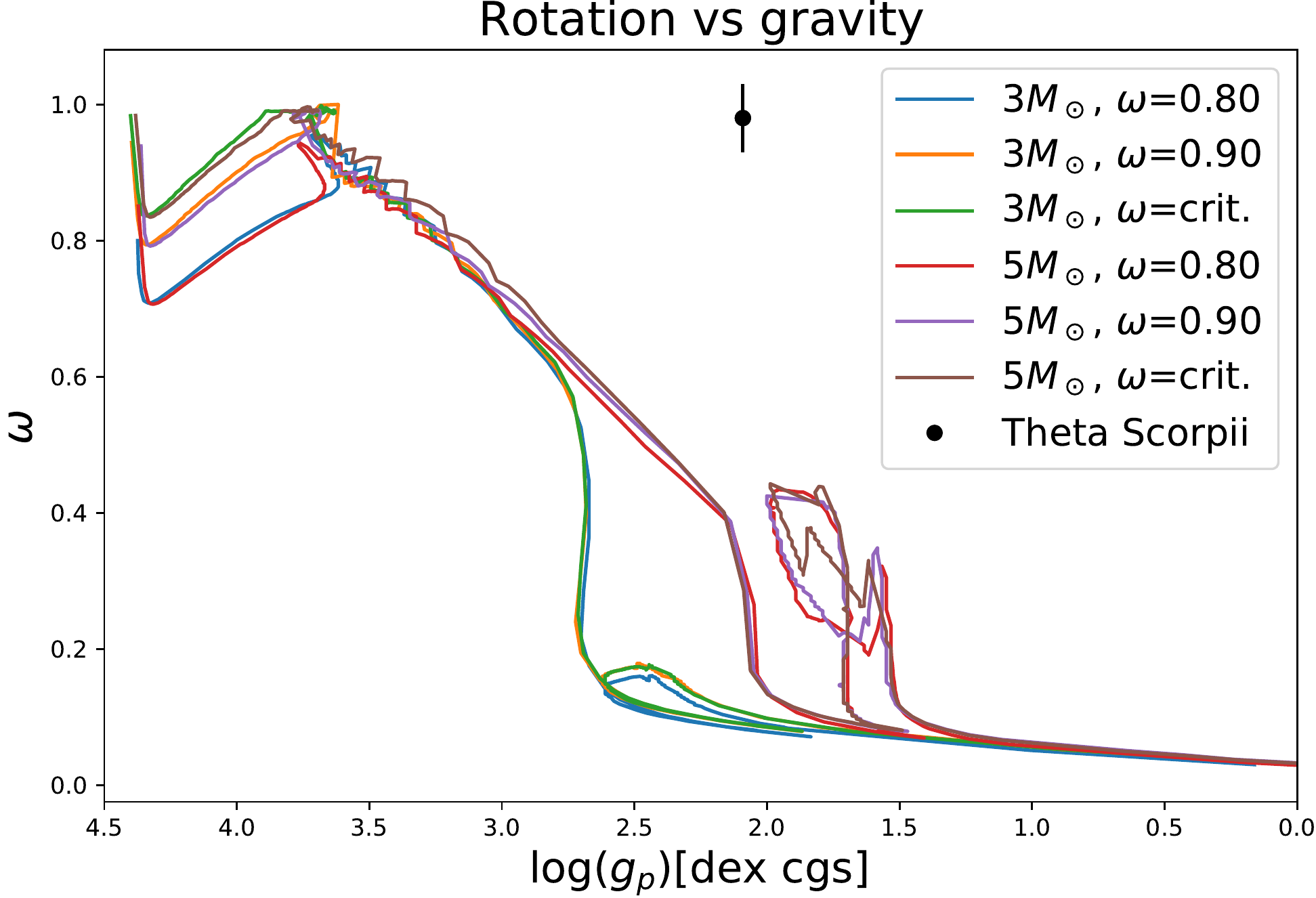}
\caption{Evolutionary tracks in the $\loggp/\omega$ plane, using solar-abundance models from \citet{Georgy}.   Tracks are identified by mass and rotation rate on the ZAMS;  evolution progresses from upper left to lower right on this plot.  Our empirical result for $\theta$~Sco~A is shown as a filled black dot with an error bar in $\omega$ (the formal uncertainty in \loggp\ is comparable to the dot size).}
 \label{fig:gvsw_tS}
\end{figure}

A major objective of this work was to measure the rotation of $\theta$~Sco~A in order to investigate its evolutionary state, particularly through the examination of the rather precisely determined gravity. A comparison with rapidly-rotating stellar-evolution models \citep{Geneva,Georgy} is shown in Figure~\ref{fig:gvsw_tS}, using
the $\loggp/\omega$ plane. Observed values
(from Table~\ref{tbl:tS_parameters}) are compared to evolutionary tracks for ZAMS masses of 3$M_\odot$ and 5$M_\odot$ (at a range of initial rotation rates), reflecting our mass estimate (3.07~$M_\odot$) as well as higher values reported in the literature \citep{Souza}. 

\begin{figure}
\includegraphics[width=8.25cm]{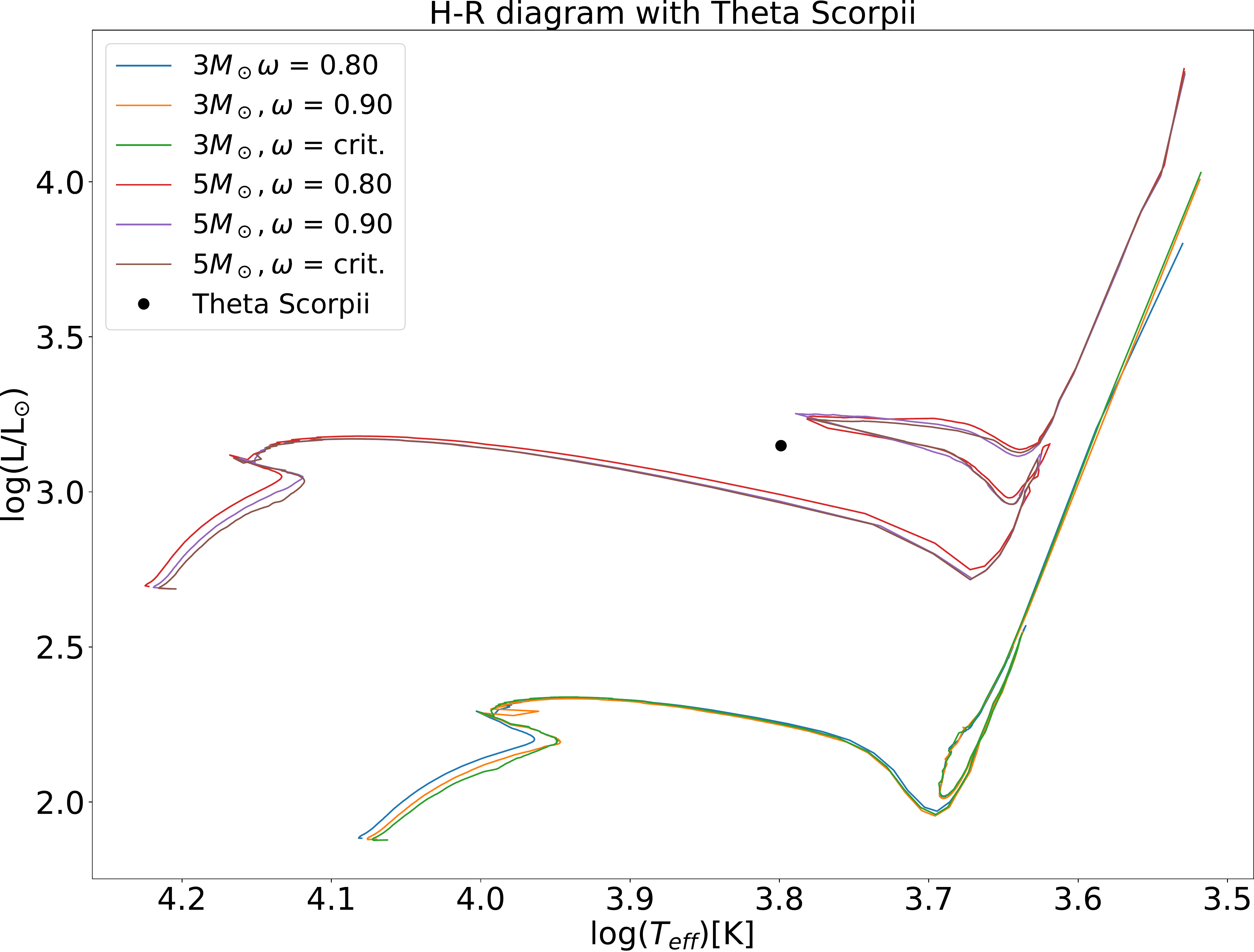}
\caption{The position of $\theta$~Sco~A in the H--R diagram, with evolutionary tracks for $3M_\odot$ and $5M_\odot$ stars.}
 \label{fig:tS_HR}
\end{figure}

It is evident from Fig,~\ref{fig:gvsw_tS} that $\theta$~Sco is rotating much faster than is consistent with any of the single-star evolutionary models. Even if it were born with critical rotation, its rotation rate would have dropped to $\omega = 0.4$ or less by the time it reached the evolutionary stage indicated by its \loggp. This conclusion holds even for the somewhat lower rotation rate obtained by \citet{Souza} from interferometry.

The conflict with single-star evolutionary tracks is further demonstrated by a standard H--R diagram (Figure~\ref{fig:tS_HR}). Based on single-star evolution one would infer that $\theta$~Sco~A is in the Hertzsprung gap and more massive than 5~\msun\ (cf.\ \citealt{Souza}). However, the mass determined from the polarization analysis is significantly lower than this.

These inconsistencies strongly suggest that $\theta$~Sco has followed a different evolutionary path to those described by
single-star models. A likely scenario is that $\theta$~Sco~A was initially a close binary system that has interacted and merged to reach its current state. The existence of such stars is not surprising. Binaries are common among massive stars \citep{duchene13} and interactions and mergers are expected to be important factors in evolution \citep{deMink2013,Sana}. It has been suggested that binary interaction may play a role in the rapid rotation of some Be stars \citep{pols91,deMink2013} and may be the predominant channel for Be star formation \citep{bodensteimer21}. However,  single-star evolutionary channels are also possible for Be stars \citep{hastings20} and so we cannot normally determine the evolutionary history in an individual case. In contrast, with an evolved star like $\theta$~Sco, the single-star evolutionary route is inconsistent with models and can be ruled out, leaving the binary route as the most likely option.

While our preferred model for the binary interaction does not involve the wider companion $\theta$~Sco~B, we cannot exclude other scenarios in which this companion was involved. Further observations to spectroscopically characterize $\theta$~Sco~B and better determine the orbit should help to clarify its involvement. If $\theta$~Sco~B was not involved in the interaction then it should be a fairly normal main sequence A star. If $\theta$~Sco~B was responsible for transferring mass and angular momentum onto $\theta$~Sco~A then it must have been initially the more massive star in the system, and would now be a stripped star. The kinematics of $\theta$~Sco show that it is not a runaway star, ruling out any interaction that involved ejection of components from the system.

We can expect to find other cases of evolved stars that are the result of binary evolution. The results presented here show that polarimetry is a useful tool for the detection and characterization of rapid rotation. Determining how common such objects are could have implications for testing predictions of binary evolution \citep{Sana} and the importance of binary interactions for stellar-population synthesis models \citep[e.g.][]{Eldridge2009,Eldridge2017}.

\section{Summary and conclusions}

Multi-wavelength, high-precision linear polarimetry of $\theta$~Sco has revealed a significant rotationally-induced stellar-polarization signal arising in the A component. The rotational polarization in this evolved star is several times higher than that previously seen in rapidly rotating main-sequence stars \citep{cotton17,bailey20b} as expected due to the lower gravity.

A reanalysis of \textit{Hipparcos} data provides the first reliable characterization of the visual-binary B~component. We find that the B component is at sub-arcsecond separation (0.538\arcsec\ in 1991 and 0.245\arcsec\ in 2021), and is $\sim$3.3 mag fainter than component A. The polarimetry, combined with additional observational constraints, permits the determination of the rotation rate and other fundamental stellar parameters of $\theta$~Sco~A at a level of precision not otherwise normally possible for single stars.

The rapid rotation we determine for $\theta$~Sco~A ($\omega \ge 0.94)$ is inconsistent with evolutionary models for single rotating stars \citep{Georgy}, which predict much slower rotation at this evolutionary stage. The mass we determine for $\theta$~Sco~A is lower than that predicted by these models. We therefore conclude that $\theta$~Sco~A is the result of a different evolutionary path, most likely interaction and eventual merger with a close binary companion.

Polarimetry can potentially be used to identify and characterize other rapidly rotating evolved stars and help to further investigate the role of binary interaction in stellar evolution.

\section*{Acknowledgements}

This paper is based in part on data obtained at Siding Spring Observatory. We acknowledge the traditional owners of the land on which the AAT stands, the Gamilaraay people, and pay our respects to elders past and present. Nicholas Borsato, Dag Evensberget, Behrooz Karamiqucham, Shannon Melrose, and Jinglin Zhao assisted with observations at the AAT. 
We thank Bob~Argyle for discussions of the visual-binary observations, and for instigating the SOAR speckle inter\-ferometry, the results of which Andrei Tokovinin generously gave us permission to quote in advance of their publication in the SOAR series (cf.\ \citealt{tokovinin21}, and references therein).   Armando Domiciano de Souza kindly commented on the likely effects of the companion on his interferometric analysis.
We acknowledge useful scientific interactions with the \textit{Polstar} development team.

\section*{Data Availability}

The new polarization data used for this project are provided in Tables~\ref{tab:observations} and~\ref{tab:controls} of the paper.



\bibliographystyle{mnras}
\bibliography{theta_sco}







\bsp	
\label{lastpage}
\end{document}